\documentclass[a4paper,12pt]{elsarticle}

\usepackage[top=0.5in,bottom=0.75in,left=0.6in,right=0.6in]{geometry}
\usepackage{graphicx}
\usepackage[colorlinks=true,linkcolor=blue,citecolor=blue,urlcolor=blue]{hyperref}
\usepackage[figurename=Fig.]{caption}
\usepackage{color}
\usepackage{float}
\usepackage{xcolor}
\usepackage{amssymb}
\usepackage{amsmath}
\usepackage{mathrsfs}
\usepackage{cases}
\usepackage{upgreek}
\usepackage{makecell}
\usepackage{pdflscape}
\usepackage{subfigure}
\usepackage{pdfpages}
\usepackage{lineno}
\usepackage{makeidx}
\usepackage{bm}
\usepackage{xstring}
\modulolinenumbers[1]
\usepackage{epstopdf}
\usepackage{lipsum}
\usepackage{slashed}
\usepackage{multicol}
\usepackage{setspace}
\usepackage{booktabs}

\usepackage[separate-uncertainty]{siunitx}

\def\degree{^\circ}

\newcommand{\doiref}[2]{\href{http://dx.doi.org/#1}{#2}}

\newcommand{\To}[1][-0.25]{\hspace{#1em}\to\hspace{#1em}}
\newcommand{\e}[1]{\times 10^{#1}}
\newcommand{\er}[3]{^{-#1}_{+#2}\times 10^{#3}}

\graphicspath{{PDF/}}

\definecolor{myblue}{RGB}{65,111,166}
\definecolor{myred}{RGB}{168,66,63}
\definecolor{mygreen}{RGB}{134,164,74}
\definecolor{mypurple}{RGB}{110,84,141}
\definecolor{myindigo}{RGB}{61,150,174}
\definecolor{myorange}{RGB}{218,129,55}
\definecolor{mylightblue}{RGB}{142,165,203}

\makeatletter
\def\ps@pprintTitle{%
  \let\@oddhead\@empty
  \let\@evenhead\@empty
  \let\@oddfoot\@empty
  \let\@evenfoot\@oddfoot
}
\makeatother

\makeatletter

\newenvironment{figurehere}
  {\def\@captype{figure}}
  {}
\makeatother

\biboptions{numbers,sort&compress}

\begin{document}


\begin{frontmatter}

\title{Decays of $B$, $B_s$ and $B_c$ to $D$-wave heavy-light mesons}

\author[]{Qiang Li}\ead{lrhit@protonmail.com}
\author[]{Tianhong Wang\corref{corauthor}}
\ead{thwang@hit.edu.cn}
\cortext[corauthor]{Corresponding author}
\author[]{Yue Jiang}\ead{jiangure@hit.edu.cn}
\author[]{Han Yuan}\ead{hanyuan@hit.edu.cn}
\author[]{Tian Zhou}\ead{tianzhou@hit.edu.cn}
\author[]{Guo-Li Wang}\ead{gl\_wang@hit.edu.cn}

\address{Harbin Institute of Technology, Harbin, 150001, P. R. China}

\begin{abstract}
We study the weak decays of $\bar{B}_{(s)}$ and $B_c$ into $D$-wave heavy-light mesons, including $J^P=2^-$ ($D_{(s)2},D'_{(s)2},B_{(s)2}, B'_{(s)2}$) and $3^-$~($D^*_{(s)3}, B^*_{(s)3}$) states. The weak decay hadronic matrix elements are achieved based on the instantaneous Bethe-Salpeter method. The branching ratios for $\bar{B}$ decays are $\mathcal{B}[\bar{B}\To D_{2}e\bar{\nu}_e] = 1.1^{-0.3}_{+0.3} \times 10^{-3}$, $\mathcal{B}[\bar{B}\To D'_2e\bar{\nu}_e]=4.1^{-0.8}_{+0.9} \times 10^{-4}$, and $\mathcal{B}[\bar{B}\To D^*_3e\bar{\nu}_e]=1.0^{-0.2}_{+0.2} \times 10^{-3}$, respectively. For semi-electronic decays of $\bar B_s$ to $D_{s2}$, $D'_{s2}$, and $D^*_{s3}$, the corresponding branching ratios are $1.7^{-0.5}_{+0.5}\times 10^{-3}$, $5.2^{-1.5}_{+1.6}\times 10^{-4}$, and $1.5^{-0.4}_{+0.4}\times 10^{-3}$, respectively. The branching ratios of semi-electronic decays of $B_c$ to $D$-wave $D$ mesons are in the order of $10^{-5}$. We also achieved the forward-backward asymmetry, angular spectra, and lepton momentum spectra. In particular the distribution of decay widths for $2^-$ states $D_2$ and $D'_2$ varying along with mixing angle are presented.
\end{abstract}

\end{frontmatter}


\section{Introduction}
The $D$-wave $D_{(s)}$ mesons have attracted lots of attention since numerous excited charmed states are discovered by {BaBar}~\cite{BaBar-2010}, and LHCb~\cite{LHCb1,LHCb2,LHCb3,LHCb4}. In 2010 {BaBar} observed four signals $D(2550)^0$, $D^*(2600)^0$, $D(2750)^0$, and $D^*(2760)^0$ for the first time~\cite{BaBar-2010}, where the last two are expected to lie in the mass region of four $D$-wave charm mesons~\cite{PRD32-1985}. Later the LHCb reported two natural parity resonances $D^*_J(2650)^0$ and $D^*_J(2760)^0$ in the $D^{*+}\pi^-$ mass spectrum and measured their angular distribution~\cite{LHCb1}. The same final states also show the presence of two unnatural parity states, $D_J(2580)^0$ and $D_J(2740)^0$. Here the natural parity denotes states with $J^P=0^+, ~1^-, ~2^+, ~3^-, \dots$ with $P=(-1)^J$, while the unnatural parity indicates series with $J^P=0^-,~1^+,~2^-,\cdots$. 

Then in May 2015, LHCb confirmed that the $D^*_J(2760)^0$ resonance has spin 1~\cite{LHCb3}. The mass and width are measured as $m[D^*_1(2760)^0]=2781\pm22$ \si{MeV} and $\Gamma[D^*_1(2760)^0]=177\pm38$ \si{MeV}, where we have combined the statistical and systematic uncertainties in quadrature for simplicity. Later LHCb determined $D^*_J(2760)^-$ to have spin-parity $3^-$ and it is interpreted as $D^*_3(2760)^-$, namely the $^3\!D_3$ $\bar{c}d$ state. The mass and width are measured as $m[D^*_3(2760)]=2798\pm10$ \si{MeV} and $\Gamma[D^*_3(2760)]=105\pm30$ \si{MeV}~\cite{LHCb4}. 

For the $D$-wave charm-strange meson, BarBar first observed the $D^*_{sJ}(2860)$~\cite{BaBar2, BaBar3}. And then LHCb's results support that $D^*_{sJ}(2860)$ is an admixture of the spin-1 and spin-3~\cite{LHCb5, LHCb6}. The measured mass and width for $D^*_{s3}$ are $2861\pm7$ and $53\pm10$ \si{MeV}, respectively. The two $D$-wave charm-strange mesons with $J=2$, namely the $2^-$ states $D_{s2}$ and $D'_{s2}$ are still undiscovered in experiment. 

Identification of these new excited charmed mesons can be found in Refs.~\cite{Col-2006,Col-2012,PRD82-2010-1,PRD82-2010-2,PRD83-2011-1,PRD83-2011-2,PRD90-2014,PRD92-2015,CPC39-2015,God-2014,PRD93-2016}. We will follow Godfrey's assignments on $D$-wave $D^{(*)}_{(s)J}$ mesons in Ref.~\cite{PRD93-2016}, where $D^*_{s3}(2860)$ is identified as $1^3\!D_3$ $c\bar{s}$; $D^*_3(2798)^0$ is identified as $1^3\!D_3(c\bar{q})$ state; $D^*_1(2760)^0$ is interpreted as $1^3\!D_1(c\bar{q})$; and the $D_J(2750)^0$ reported by BaBar and ${D_J(2740)^0}$ reported by LHCb are identified as the same state with $1D_2(c\bar{q})$, where $q$ denotes a light quark $u$ or $d$. 

These $D$-wave excited states still need more experimental data to be discovered or confirmed. The identification and spin-parity assignments in above literature are just tentative. As the LHC accumulates more and more data, the study of these $D$-wave charm and charm-strange mesons in the weak decay of $B_{(s)}$ and $B_c$ meson becomes necessary and important. The properties of $D^{(*)}_{(s)J}$ in $B_{(s)}$ and $B_c$ decays would be helpful in identification of these excited $D_{(s)}$ mesons. The semi-leptonic decays of $B_{(s)}$ into $D$-wave charmed mesons have been studied by QCD sum rules~\cite{PLB478-2000, PRD79-2009, Gan-2015} and constituent quark models in the framework of heavy quark effective theory~(HQET)~\cite{PTP91-1994, PRD54-1996}. Most of previous work is based on the HQET. The systematic studies on weak decays of $\bar B_{(s)}$  into $D$-wave $D_{(s)2}$, $D'_{(s)2}$ and $D^*_{(s)3}$ are still quite less while all these $D$-wave charmed mesons are hopefully to be detected in the near future experiments. On the other hand, in 2012 the BaBar Collaboration reported the ratio of $\mathcal{B}(\bar B\To D^{(*)}\tau^-\bar \nu_\tau)$ relative to $\mathcal{B}(\bar B\To D^{(*)}e^-\bar \nu_e)$, which exceed the standard model expectation by $2\sigma~(2.7\sigma)$~\cite{BaBar-2012} and may hint the new physics. We also noticed that in the very recently Belle measurement\,\cite{Belle-2016}, the experimental results on this quantity are consistent with the theoretical predictions\,\cite{Fajfer-2012,Tanaka-2013,Xiao-2015} in the framework of the Standard Model. Anyway, it is still necessary and helpful to investigate these ratios for $\bar B_{(s)}$ and $B_c$ decays into higher excited $D_{(s)}$ mesons. 

In this work we will concentrate on the semi-leptonic and non-leptonic decays of $\bar{B}$ ($\bar{B}_s, B_c$) into $D$-wave $D$ ($D_s$) meson, including $2^- ~(D_{(s)2}, D'_{(s)2})$ and $3^-~(D^*_{(s)3})$ states. For completeness, the weak decays of $B_c$ to $D$-wave bottomed mesons are also studied. We use the Instantaneous Bethe-Salpeter equation (IBS)~\cite{PR87-1952} to get the hadronic transition form factors. BS equation~\cite{PR84-1951} is the relativistic two-body bound states formula. Based on our previous studies~\cite{NPP39-2012,JPG40-2013,JHEP09-2015,QLi-2016}, the relativistic corrections for transitions of higher excited states are larger and more important than that for the ground states, so the relativistic method is more reliable for the processes involved the high excited states.  In the instantaneous approximation of the interaction kernel, we can achieve the Salpeter equation. The Salpeter method has been widely used to deal with heavy mesons' decay constants calculation~\cite{PLB633-2006, Guo-2007}, annihilation rate~\cite{PLB674-2009, JHEP03-2013}, and hadronic transition~\cite{NPP39-2012,JPG40-2013,JHEP09-2015,QLi-2016}.

This paper is organized as follows: first we present the general formalism of semi-leptonic and non-leptonic decay for $\bar B_{(s)}$ meson, including decay width, forward-backward asymmetry, and lepton spectra. In~\autoref{Sec-3} we compute the form factors in hadronic transition by Salpeter method.  In~\autoref{Sec-4} we give the numerical results and discussions. Finally we give a short summary of this work.

\section{Formalism of semi-leptonic and non-leptonic decays}\label{Sec-2}
In this section, firstly we will derive the formalism of transition amplitudes for $\bar{B}_{(s)}$ to $D$-wave heavy-light mesons. Then the formalisms of interested observables are presented. We will take the $\bar B\To D^{(*)}_{J}$ transition as an example to show the calculation details, while results for transition of $B_s$ and $B_c$ will be given directly.
\subsection{Semi-leptonic decay amplitude}
\begin{figure}[h]
\centering
\includegraphics[width=0.55\textwidth]{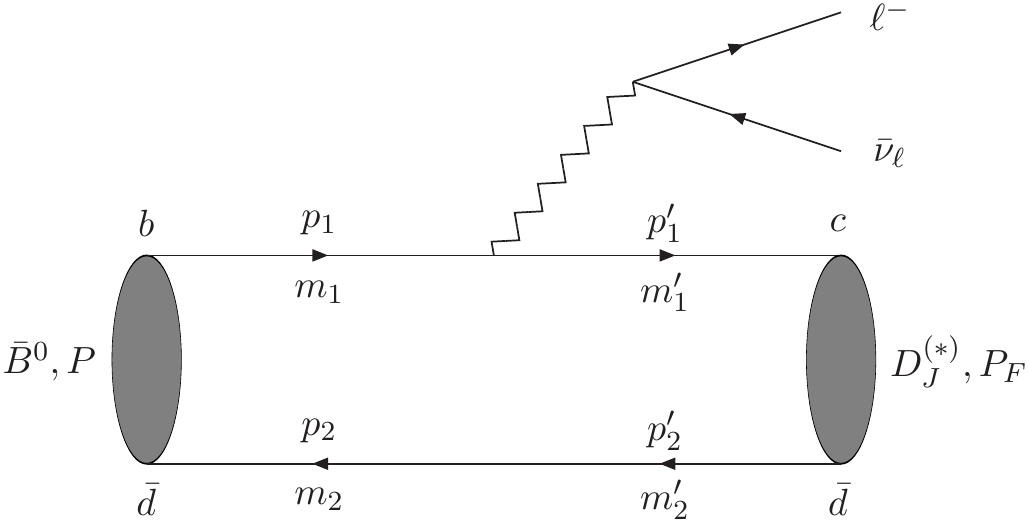}
\caption{Feynman diagram for semi-leptonic decay of $\bar{B}$ to $D^{(*)}_{J}$ $(J=2, 3)$.  $m_i(m'_i)$ and $p_i(p_i^\prime)$ are the constituent quark mass and momentum for initial (final) state, respectively.}\label{Fig-semi}
\end{figure}

The Feynman diagram responsible for $\bar{B}$ semi-leptonic decay is showed in~\autoref{Fig-semi}, where we use $P$ and $P_F$ to denote the momenta of $\bar{B}$ and $D^{(*)}_{J}$ respectively.
The transition amplitude $\mathcal{A}$ for the process $\bar B\To D^{(*)}_J\ell\bar\nu$ can be written directly as
\begin{equation}\label{Amp}
\mathcal{A}=\frac{G_{\!F}}{\sqrt{2}}V_{cb} l^\mu \langle D^{(*)}_{J}|J_{\mu}|\bar{B}\rangle.
\end{equation} 
In above equation, $G_{\!F}$ denotes the Fermi weak coupling constant; $V_{cb}$ is the Cabibbo-Kobayashi-Maskawa matrix element; the lepton matrix element $l^\mu$ reads 
\begin{gather}
l^\mu=\bar{u}(p_\ell)\Gamma^\mu v(p_\nu),
\end{gather} 
where $\ell$~($\bar{\nu}_\ell$) denotes the charged lepton (anti-neutrino), and $p_\ell$($p_\nu$) denotes the corresponding momentum, and the definition $\Gamma^\mu=\gamma^\mu(1-\gamma^5)$ is used;~$\langle D^{(*)}_{J}|J_{\mu}|\bar{B}\rangle$ is the hadronic transition element, where $J_\mu=\bar{c}\Gamma_\mu b$ is the weak current.

We use $\mathcal{M}^\mu$ to denote hadronic transition element  $\langle D^{(*)}_J|J^{\mu}|\bar{B}\rangle$, which can be described with form factors. The general form of the hadronic matrix element depends on the total angular momentum $J$ of the final meson. For $D_2~(D'_2)$ and $D^*_3$ the form factors are defined as 
\begin{equation} \label{form}
\mathcal{M}^\mu=
\begin{cases} 
e_{\alpha\beta}P^{\alpha}(s_1P^{\beta}P^{\mu}+s_2P^{\beta}P_F^{\mu}+s_3g^{\beta\mu}+\mathrm{i}s_4\epsilon^{\mu\beta PP_F})   & \text{if}~J=2,\\
e_{\alpha\beta\gamma}P^{\alpha}P^{\beta}(h_1P^{\gamma}P^{\mu}+h_2P^{\gamma}P_F^{\mu}+h_3
g^{\gamma\mu}+\mathrm{i}h_4\epsilon^{\mu \gamma PP_F})    & \text{if}~J=3.
\end{cases}
\end{equation}
In above equation, we used the definition $\epsilon_{\mu \nu P P_F}=\epsilon_{\mu \nu \alpha \beta }P^{\alpha} P^{\beta}_F$ where $\epsilon_{\mu \nu \alpha \beta}$ is the totally antisymmetric Levi-Civita tensor; $g^{\mu\nu}$ is the Minkowski metric tensor; $e_{\alpha\beta}$ and  $e_{\alpha\beta\gamma}$ are the polarization tensor for $J=2$ and 3 mesons, respectively, which are completely symmetric; $s_i$ and $h_i~(i=1,2,3,4)$ are the form factors for $J=2$ and 3, respectively.
To state it more clearly, we will use $s_i$, $t_i$, and $h_i$ to denote the form factors for transitions $\bar B$ to $D_2$, $D'_2$, and $D^*_3$, respectively.
$S_i$, $T_i$, and $H_i$ are used to represent the form factors of $B^-_c$ to $\bar D_2$, $ D'_2$, and $ D^*_3$, respectively. The definition forms are the same with that for transition $\bar B\To D^{(*)}_J$, just $s_i$ is replaced by $S_i$, $t_i$ by $T_i$, and $h_i$ by $H_i$. For $\bar B_s$ decays, the corresponding form factor behaviors are very similar to $\bar B$ decays. The detailed calculations of these form factors will be given in next section. 

After summing the polarization of all the final states, including the charged lepton, anti-neutrino and the final $D^{(*)}_J$, we obtain
\begin{equation} \label{Amp2}
\overline{|\mathcal{A}|^2} =\frac{G_F^2}{2} |V_{cb}|^2 L^{\mu \nu} H_{\mu\nu} ,
\end{equation}
where the lepton tensor $L^{\mu \nu}$ has the following form
\begin{equation} \label{T-L}
L^{\mu \nu}= 8(p_\ell^{\mu}p_\nu^{\nu}+p_\nu^{\mu}p_\ell^{\nu}-p_\ell \! \cdot \! p_\nu g^{\mu \nu}-\mathrm{i}\epsilon^{\mu \nu p_\ell p_\nu}).
\end{equation}
$H_{\mu \nu}$ is the hadronic tensor describing the propagator-meson interaction vertex, which depends on $P$, $P_F$ and the corresponding form factors. It can be written as
\begin{gather} \label{eq-HF}
H_{\mu \nu}=\sum_{s=-J}^{J} \mathcal{M}_\mu^{(s)}\mathcal{M}^{*(s)}_\nu= N_{1}P_{\mu}P_{\nu}+N_2(P_{\mu}{P_F}_\nu+P_{\nu}{P_F}_\mu)+N_4P_{F\mu}P_{F\nu}+
N_5g_{\mu \nu}+\mathrm{i} N_6 \epsilon_{\mu \nu P P_F},
\end{gather}
where the summation is over the polarization of final $D^{(*)}_J$ meson; $N_i$ is related to the form factors $s_i$ for $D_2$, $t_i$ for $D'_2$ or $h_i$ for $D^*_3$. The detailed expressions for $N_i$ can be found in appendix~\ref{Ni}.

\subsection{Non-leptonic decay amplitude}
\begin{figure}[ht]
\centering
\includegraphics[width = 0.5\textwidth]{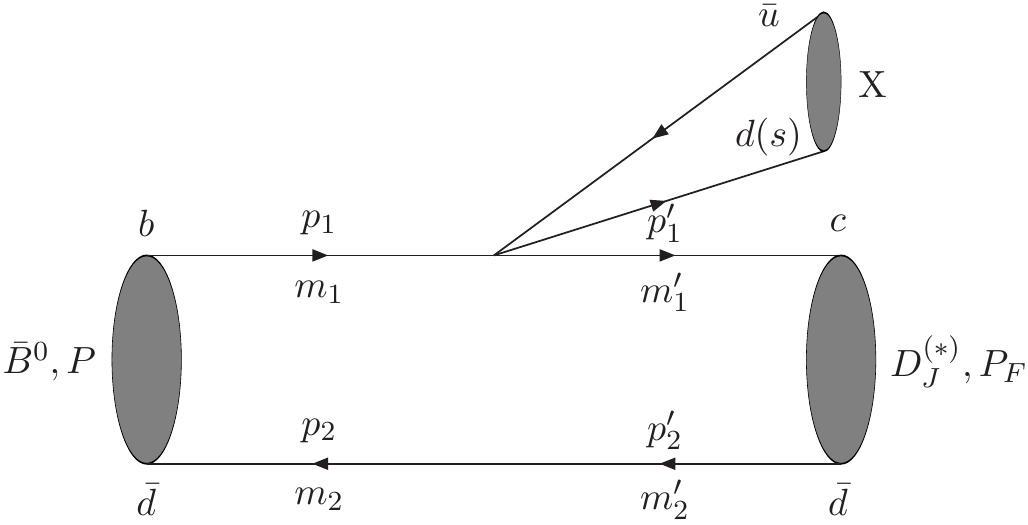}
\caption{The Feynman diagram of the nonleptonic decay of $\bar B_{(s)}$ meson into $D$-wave charmed meson. $X$ denotes a light meson.}\label{bd-non}
\end{figure}

The Feynman diagram for the non-leptonic decay of $\bar B$ to $D^{(*)}_{J}$ and a light meson $X$ is showed in \autoref{bd-non}. As a preliminary study for non-leptonic decays of $\bar B$ to $D$-wave $D$ mesons, we will work in the framework of naive factorization approximation~\cite{Fak-1978, Cab-1978,Bau-1987,Ali-1998}, which has been widely used in heavy mesons' weak decays~\cite{cch1,Col-2000,PRD73-2006,PRD74-2006,Faustov-2013}. The factorization assumption is expected to hold for process that involves a heavy meson and a light meson, provided the light meson is energetic~\cite{Dugan-1991,Ben-1999,Keu-2001}. Also we only consider the processes when the light meson $X$ is $\pi$, $\rho$, $K$, or $K^*$.
In the naive factorization approximation, the decay amplitude can be factorized as the product of two parts, the hadronic transition matrix element and an annihilation matrix element. Then we can write the non-leptonic decay amplitude as 
\begin{equation} \label{Amp-non}
\mathcal{A}\big [\bar B\To D^{(*)}_{J}X \big ] \simeq \frac{G_F}{\sqrt{2}} V_{cb} V_{uq} a_1(\mu) \langle D^{(*)}_{J}|J_\mu|\bar B \rangle \langle X|(\bar qu )_{V-A}|0\rangle,
\end{equation}
where we have used the definition $(\bar q u)_{V-A}=\bar{q} \Gamma^\mu  u$ and $q$ denotes a $d$ or $s$ quark field; $V_{uq}$ denotes the corresponding CKM matrix element; $a_1=c_1+\frac{1}{N_c}c_2$, where $N_c=3$ is the number of colors. For $b$ decays, we take $\mu=m_b$, and $a_1=1.14$, $a_2=-0.2$~\cite{PRD73-2006} are used in this work. 
The annihilation matrix element can be expressed by decay constant as 
\begin{equation}
\langle X|(\bar qu )_{V-A}|0\rangle=
\begin{cases}
\text{i} P^{\mu}_X f_P~~~~~&X{\rm~is~a~pseudoscalar~meson}~(\pi, K),\\
 e^{\mu}M_X f_V       &{X\rm ~is~a~vector~meson}~(\rho, K^*).
\end{cases}
\end{equation}
$M_X$, $P_X$ are the mass and momentum of $X$ meson, respectively; the meson polarization vector $e^\mu$ satisfies $e_\mu P^\mu_X=0$ and the completeness relation is given by $\sum_s e^\mu_{(s)} e^\nu_{(s)}=\frac{P_X^\mu P_X^\nu}{M_X^2}-g^{\mu\nu}$, where $s$ denotes the polarization state; $f_P$ and $f_V$ are the corresponding decay constants.

Then the $\overline{|\mathcal{A}|^2}$ can be expressed by hadronic tensor $H_{\mu\nu}$, which is just the same with that in the corresponding semi-leptonic decays, and light meson tensor $X^{\mu\nu}$ as
\begin{equation}
\overline{|\mathcal{A}|^2}=\frac{G_F^2}{2}|V_{cb}|^2|V_{uq}|^2a_1^2H_{\mu\nu}X^{\mu\nu},
\end{equation}
where $X^{\mu\nu}$ has the following expression
\begin{equation}
X^{\mu\nu}=\langle X|(\bar q\Gamma^\mu u )|0\rangle\langle X|\bar q\Gamma^\nu u|0\rangle^*=
\begin{cases}
P^{\mu}_X P^{\nu}_X f^2_P 					    &X{\rm~is~a~pseudoscalar~meson},\\
(P^{\mu}_X P^{\nu}_X- M^2_Xg^{\mu\nu}) f^2_V        &X{\rm ~is~a~vector~meson}.
\end{cases}
\end{equation}

\subsection{Several observables}
One of the interested quantity in $\bar B$ semi-leptonic decay is the angular distribution of the decay width $\Gamma$, which can be described as
\begin{equation}
\frac{\text{d} \Gamma}{ \mathrm{d}\cos\theta} = \int \frac{1}{(2\uppi)^3} \frac{|\bm{p}^*_\ell||\bm{p}_F^*|}{16M^3} \overline{|\mathcal{A}|^2} \text{d} m^2_{\ell\nu},
\end{equation}
where $M$ is the initial $\bar B$ mass; $m^2_{\ell\nu}=(p_\ell+p_\nu)^2$ is the invariant mass square of $\ell$ and $\bar\nu$; $\bm{p}^*_\ell$ and $\bm{p}_F^*$ are the three momenta of $\ell$ and $D^{(*)}_{J}$ in the $\ell\bar\nu$ rest frame, respectively; $\theta$ is the angle between  $\bm{p}^*_\ell$ and $\bm{p}_F^*$;  $|\bm{p}^*_{\ell}|=\frac{1}{2m_{\ell\nu}}\lambda^{\frac{1}{2}}(m^2_{\ell\nu}, M_{\ell}^2, M^2_{\nu})$ and $|\bm{p}_F^*|=\frac{1}{2m_{\ell\nu}}\lambda^{\frac{1}{2}}(m^2_{\ell\nu}, M^2, M^2_F)$, where we have used the K\"all\'en function $\lambda(a,b,c)=(a^2+b^2+c^2-2ab-2bc-2ac)$, $M_{\ell}$ and $M_{\nu}$ are the lepton mass and anti-neutrino mass, respectively. Another quantity we are interested is the forward-backward asymmetry $A_{FB}$, which is defined as 
\begin{equation}
A_{FB} = \frac{\Gamma_{\cos\theta>0}-\Gamma_{\cos\theta<0}}{\Gamma_{\cos\theta>0}+\Gamma_{\cos\theta<0}}.
\end{equation}
The decay width varying along with charged lepton 3-momentum $|\bm{p}_\ell|$ is given by
\begin{equation}
\frac{\text{d} \Gamma}{\text{d} |\bm{p}_\ell|} =\int \frac{1}{(2\uppi)^3} \frac{|\bm{p}_\ell|} {16M^2 E_\ell} \overline{|\mathcal{A}|^2} \text{d} m^2_{\ell\nu} ,
\end{equation}
where $E_\ell$ denotes the charged lepton energy in the rest frame of initial state meson.

The non-leptonic decay width of the $\bar{B}$ meson is given by 
\begin{equation} \label{eq-width2}
\Gamma=\frac{|\bm{p}|}{8\uppi M^2} \overline{|\mathcal{A}|^2},
\end{equation}
where $\bm{p}$ represents the 3-momentum of the final $D^{(*)}_{J}$ in $\bar B$ rest frame, which is expressed as $|\bm{p}|=\frac{1}{2M}\lambda^{\frac{1}{2}}(M^2, M^2_X, M^2_F)$.

\section{Hadronic transition matrix element}\label{Sec-3}
The hadronic transition matrix element $\langle D^{(*)}_{J}|J^{\mu}|\bar{B}\rangle$ plays an key role in the calculations of $\bar B$ semi-leptonic and non-leptonic decays. In this section we will give details to calculate the hadronic transition matrix element by Bethe-Salpeter method in the framework of constituent quark model.

\subsection{Formalism of hadronic transition matrix element with Bethe-Salpeter method}

According to the Mandelstam formalism~\cite{Man-1955}, the hadronic transition amplitude $\mathcal{M}^\mu$ can be written by Beter-Salpeter~(BS) wave function as
\begin{equation} \label{M-0}
\mathcal{M}^\mu=-\mathrm{i} \int \frac{\mathrm{d}^4 q \mathrm{d}^4 q'}{(2 \uppi)^4} \mathrm{Tr} \big[ \bar{\Psi}_D(q',P_F) \Gamma^{\mu} \Psi_B(q,P) (m_2+\slashed {p}_2) \delta^{4}(p_2-p'_2)\big],
\end{equation}
where $\Psi_B(q,P)$ and $\Psi_D(q',P_F)$ are the BS wave functions of the $\bar B$ meson and the final $D^{(*)}_J$, respectively; $\bar\Psi$ is defined as $\gamma^0\Psi^\dagger\gamma^0$; $q$ and $q'$ are respectively the inner relative momenta of $\bar B$ and $D^{(*)}_J$ system, which are related to the quark (anti-quark) momentum $p_1^{(\prime)}$ ($p_2^{(\prime)}$) by $p_i=\alpha_i P+(-1)^{(i+1)}q$ and $p'_i=\alpha^\prime_i P_F+(-1)^{(i+1)}q^\prime$ ($i=1, 2$). And here we defined the symbols $\alpha_i=\frac{m_i}{m_1+m_2}$ and $\alpha^\prime_i=\frac{m^\prime_i}{m'_1+m'_2}$, where $m_i$ and $m'_i$ are masses of the constituent quarks in the initial and final bound states, respectively (see \autoref{Fig-semi}). Here in $\bar B$ decays we have $m_1=m_b$, $m'_1=m_c$, $m_2=m'_2=m_d$. As there is a delta function in above equation, the relative momenta $q$ and $q^\prime$ are related by $q '=q-(\alpha_2 P-\alpha '_2 P_F)$.

In the instantaneous approximation~\cite{PR87-1952}, the inner interaction kernel between quark and anti-quark in bound state is independent of the time component $q_P(=q\cdot P)$ of $q$. By performing the contour integral on $q_P$ and then we can express the hadronic transition amplitude as~\cite{QLi-2016} 
\begin{equation}\label{M}
\mathcal{M}^\mu = \int \frac{ \text{d}^3 {q_\perp}}{(2\uppi)^3} \text{Tr}   \bigg [  \frac{\slashed{P}}{M} \bar{\psi}_{D}(q'_\perp) \Gamma^\mu \psi_B(q_\perp)  \bigg  ],
\end{equation}
where we have used the definitions $q_\perp\equiv q-\frac{P\cdot q}{M^2}P$ and $q'_\perp\equiv q'-\frac{P\cdot q'}{M^2}P$. Here $\psi$ denotes the 3-dimensional positive Salpeter wave function~(see appendix \ref{Salpeter-EQ}). $\psi_B$ and $\psi_D$ denote the positive Salpeter wave functions for $\bar B$ and $D^{(*)}_J$, respectively, and $\bar \psi_D$ is defined $\gamma^0 \psi_D^\dag \gamma^0$.

The positive Salpeter wave function for ${^1\!S_0}(0^-)$ state can be written as~\cite{PLB584-2004}
\begin{equation} \label{wave-1s0}
\psi_B(^1\!S_0)=\bigg [ A_1+A_2 \frac{\slashed P}{M} +A_3 \frac{\slashed q_{\perp}}{M}+A_4\frac{\slashed P \slashed q_{\perp}}{M^2}  \bigg ]\gamma^5,
\end{equation}
where we have the following constraint conditions,
\begin{equation}\label{par-1s0}
\begin{aligned}
A_1=& \frac{M}{2}\bigg[\frac{\omega_1+\omega_2}{m_1+m_2}k_1+k_2 \bigg],~
&A_3=&-\frac{M(\omega_1-\omega_2)}{m_1\omega_2+m_2\omega_1}A_1,	\\
A_2=&\frac{M}{2}\bigg[k_1+\frac{m_1+m_2}{\omega_1+\omega_2}k_2 \bigg],~
&A_4=&-\frac{M(m_1+m_2)}{m_1\omega_2+m_2\omega_1}A_1.
\end{aligned}
\end{equation}
The definition $\omega_i\equiv \sqrt{m_i^2-q_\perp^2}~(i=1,2)$ is used. The derivation of Eq.(\ref{wave-1s0}) and (\ref{par-1s0}) can be found in appendix \ref{Salpeter-EQ}.  So there are only two undetermined wave function $k_1$ and $k_2$ here, which are the functions of $q_\perp$.
The positive Salpeter wave function for $3^-(^3\!D_3)$ state with unequal mass of quark and anti-quark has the following forms~\cite{3D23-wave}
\begin{equation} \label{wave-3d3}
\begin{aligned}
\psi_D(^3\!D_3)&=e_{\mu \nu \alpha} q^{\prime\nu}_{\perp} q_{\perp}^{\prime\alpha} \biggl [ q^{\prime\mu}_{\perp}( n_1+n_2\frac{\slashed P_F}{M_F}+n_3 \frac{\slashed q'_{\perp}}{M_F} +n_4\frac{\slashed P_F \slashed q'_{\perp}}{M_F^2})
+ \gamma^\mu ( n_5 M_F + n_6 \slashed P_F) \\
&\quad+n_7 (\gamma^\mu \slashed{q'}-q^{\prime\mu})+ n_8  \frac{(\gamma^\mu \slashed P_F \slashed q'_{\perp} +  \slashed P_F q^{\prime\mu}_{\perp})}{M_F}\bigg].
\end{aligned}
\end{equation}
In above equation $n_i~(i=1,2,\cdots,8)$ can be expressed with 4 wave functions $u_i~(i=3,4,5,6)$ as below,
\begin{equation}\label{par-3d3}
\begin{aligned}
n_1=& \frac{\big[(\omega'_1+\omega'_2)(q_\perp^{\prime2}u_3+M_F^2u_5)+(m'_1+m'_2)(q_\perp^{\prime2}u_4-M_F^2 u_6)\big]}{2M_F(m'_1 \omega'_2+m'_2 \omega'_1)},\\
n_2=& \frac{\big[(m'_1-m'_2)(q_\perp^{\prime2}u_3+M_F^2u_5)+(\omega'_1-\omega'_2)(q_\perp^{\prime2}u_4-M_F^2 u_6)\big]}{2M_F(m'_1 \omega'_2+m'_2 \omega'_1)},\\
n_3=&\frac{1}{2}\biggl[u_3+\frac{m'_2+m'_2}{\omega'_1+\omega'_2}u_4-\frac{2M_F^2}{m'_1\omega'_2+m'_2\omega'_1}u_6 \biggl],\\
n_4=&\frac{1}{2}\biggl[u_4+\frac{\omega'_1+\omega'_2}{m'_1+m'_2}u_3-\frac{2M_F^2}{m'_1\omega'_2+m'_2\omega'_1}u_5 \biggl],\\
n_5=& \frac{1}{2}\biggl[u_5-\frac{\omega'_1+\omega'_2}{m'_1+m'_2}u_6 \biggl],\quad
n_6= \frac{1}{2}\biggl[u_6-\frac{m'_1+m'_2}{\omega'_1+\omega'_2}u_5\biggl],\\
n_7=& \frac{M_F(\omega'_1-\omega'_2)}{(m'_1\omega'_2+m'_2\omega'_1)}n_5,~~~~\quad
n_8= \frac{M_F(\omega'_1+\omega'_2)}{(m'_1\omega'_2+m'_2\omega'_1)}n_6.
\end{aligned}
\end{equation}
In above Salpeter positive wave functions $\psi_B$ and $\psi_D$, the undetermined wave functions $k_1$, $k_2$ for $0^-$ and $u_i~(i=3,4,5,6)$ for $3^-$ can be achieved by solving the full Salpeter equations numerically~(see appendix \ref{Salpeter-EQ}).  The positive Salpeter wave functions for $^1\!D_2$~\cite{JHEP03-2013}, and $^3\!D_2$~\cite{3D23-wave} states can be seen in appendix~\ref{sec-wave}. $e^{\mu\nu\alpha}$ is the symmetric polarization tensor for spin-3 and satisfies the following  relations~\cite{PRD43-1991}
\begin{gather}\label{Polarization3}
e^{\mu\nu\alpha} g_{\mu \nu}=0,\quad
e^{\mu\nu\alpha} P_{F\mu}=0,\\
\sum_s e^{abc}_{(s)}e^{\mu\nu\alpha}_{(s)}=\frac{1}{6}\Omega_1^{abc;\mu\nu\alpha}-\frac{1}{15}\Omega_2^{abc;\mu\nu\alpha},
\end{gather}
where
\begin{equation}\label{eq-Polar31}
\begin{aligned}
\Omega_1^{abc;\mu\nu\alpha}&=g_\perp^{a\mu}g_\perp^{b\nu}g_\perp^{c\alpha}+
g_\perp^{a\mu}g_\perp^{b\alpha}g_\perp^{c\nu}+
g_\perp^{a\nu}g_\perp^{b\mu}g_\perp^{c\alpha}
+g_\perp^{a\nu}g_\perp^{b\alpha}g_\perp^{c\mu}+
g_\perp^{a\alpha}g_\perp^{b\mu}g_\perp^{c\nu}+g_\perp^{a\alpha}g_\perp^{b\nu}g_\perp^{c\mu},\\
\Omega_2^{abc;\mu\nu\alpha}&=g_\perp^{ab}g_\perp^{c\mu}g_\perp^{\nu\alpha}+
g_\perp^{ab}g_\perp^{c\nu}g_\perp^{\mu\alpha}+
g_\perp^{ab}g_\perp^{c\alpha}g_\perp^{\mu\nu}
+g_\perp^{ac}g_\perp^{b\mu}g_\perp^{\nu\alpha}+
g_\perp^{ac}g_\perp^{b\nu}g_\perp^{\mu\alpha}
+g_\perp^{ac}g_\perp^{b\alpha}g_\perp^{\mu\nu}\\
&+g_\perp^{bc}g_\perp^{a\mu}g_\perp^{\nu\alpha}+
g_\perp^{bc}g_\perp^{a\nu}g_\perp^{\mu\alpha}+g_\perp^{bc}g_\perp^{a\alpha}g_\perp^{\mu\nu}.\\
\end{aligned}
\end{equation}
and we have used the definition $g_\perp^{\mu\nu}=-g^{\mu\nu}+\frac{P_F^\mu P_F^\nu}{M_F^2}$.
 
Inserting the initial $\bar B$ wave function $\psi_B(^1\!S_0)$ (Eq.~\ref{wave-1s0}) and final $D^{*}_3$ wave function $\psi_D(^3\!D_3)$ (Eq.~\ref{wave-3d3}) into the hadronic transition amplitude Eq.~(\ref{M}), after calculating the trace and performing the integral in Eq.~(\ref{M}) we achieve the form factors $h_i$ for $\bar{B}\To D^{*}_3$ transition defined in Eq.~(\ref{form}). When performing the integral over $\bm{q}$ in the rest frame of the initial meson, the following formulas are used.
\begin{gather*}
\int \frac{\mathrm{d}^3\bm q}{(2\pi)^3}q^{\mu}_\perp       = ~C_1P_{F\perp}^{\mu},\\
\int \frac{\mathrm{d}^3\bm q}{(2\pi)^3}q^{\mu}_\perp q^{\nu}_\perp= ~C_{21}P_{F\perp}^{\mu}P_{F\perp}^{\nu}+C_{22}g_T^{\mu \nu},\\
\int \frac{\mathrm{d}^3\bm q}{(2\pi)^3}q^{\mu}_\perp q^{\nu}_\perp q^{\alpha}_\perp=  ~C_{31}P_{F\perp}^{\mu}P_{F\perp}^{\nu}P_{F\perp}^{\alpha}
+ C_{32}(g_T^{\mu \nu}P_{F\perp}^{\alpha}+g_T^{\mu \alpha}P_{F\perp}^{\nu}+g_T^{\alpha \nu}P_{F\perp}^{\mu}),\\
\int \frac{\mathrm{d}^3\bm q}{(2\pi)^3}q^{\mu}_\perp q^{\nu}_\perp q^{\alpha}_\perp q^{\beta}_\perp=  ~C_{41}P_{F\perp}^{\mu}P_{F\perp}^{\nu}P_{F\perp}^{\alpha}P_{F\perp}^{\beta}+
C_{42}(g_T^{\mu \nu}P_{F\perp}^{\alpha}P_{F\perp}^{\beta}+g_T^{\mu \alpha}P_{F\perp}^{\nu}P_{F\perp}^{\beta}+
g_T^{\alpha \nu}P_{F\perp}^{\mu}P_{F\perp}^{\beta}+\\  g_T^{\alpha \beta}P_{F\perp}^{\mu}P_{F\perp}^{\nu}+g_T^{\beta \nu}P_{F\perp}^{\mu}P_{F\perp}^{\alpha}+g_T^{\beta \mu}P_{F\perp}^{\nu}P_{F\perp}^{\alpha})+
C_{43}(g_T^{\alpha \beta}g_T^{\mu \nu}+g_T^{\alpha \nu}g_T^{\mu \beta}+g_T^{\alpha \mu}g_T^{\beta \nu}),
\end{gather*}
where $g_T^{\mu\nu}$ are defined as $(g^{\mu\nu}-\frac{P^\mu P^\nu}{P^2})$ and $P^\mu_{F\perp}=(P_F^\mu-\frac{P_F\cdot P}{M^2}P^\mu)$. From above equations we can easily obtain the following expressions of $C_i$,
\begin{equation}\label{eq-Ci}
\left \{
\begin{aligned}
C_{1~}=&~ |\bm q| \cos \eta,                                  &\qquad            C_{21}=&~\frac{1}{2}|\bm q|^2(3\cos^2 \eta-1),\\
C_{22}=&~\frac{1}{2}|\bm q|^2(\cos^2 \eta-1),                 &\qquad            C_{31}=&~\frac{1}{2}|\bm q|^3(5\cos^3 \eta-3 \cos \eta),\\
C_{32}=&~\frac{1}{2}|\bm q|^3(\cos^3 \eta- \cos \eta),      &\qquad            C_{41}=&~\frac{1}{8}|\bm q|^4(35\cos^4 \eta-30\cos^2 \eta+3),\\
C_{42}=&~\frac{1}{8}|\bm q|^4(5\cos^4 \eta-6\cos^2 \eta+1), &\qquad            C_{43}=&~\frac{1}{8}|\bm q|^4(\cos^4 \eta-2\cos^2 \eta+1),
\end{aligned}
\right .
\end{equation}
where $\eta$ is the angle between $\bm{q}$ and $\bm{P}_F$.

The physical $2^-$ $D$-wave states $D_2$ and $D'_{2}$ are the mixing states of $^3\!D_2$ and ${^1\!D_2}$ states, whose wave functions are what we solve directly from the full Salpeter equations. 
Here we will follow Ref.~\cite{Matsuki-2010} and Ref.~\cite{Fau-2010}, where the mixing form for $D$-wave states is defined with the mixing angle $\alpha$ as
\begin{equation}\label{def-D2}
\begin{aligned}
|D_2\rangle  &= +\cos \alpha~ |^1\!D_2\rangle + \sin \alpha ~|^3\!D_2\rangle, \\
|D'_2\rangle &= -\sin \alpha ~|^1\!D_2\rangle + \cos \alpha~ |^3\!D_2\rangle.
\end{aligned}
\end{equation}
In the heavy quark limit ($m_Q\To\infty$), the $D$ mesons are described in the $|J,j_\ell\rangle$ basis, where $m_Q$ denotes the heavy quark mass and $j_\ell$ denotes the total angular momentum of the light quark. The relations between $|J,j_\ell\rangle$ and $|J,S\rangle$ for $L=2$ are showed by 
\begin{gather}
\begin{bmatrix} |2,5/2\rangle \\ |2, 3/2\rangle \end{bmatrix}=\frac{1}{\sqrt{5}}\begin{bmatrix}\sqrt{2+1} & \sqrt{2}\\ -\sqrt{2} & \sqrt{2+1}\end{bmatrix} \begin{bmatrix} |^1\!D_2\rangle \\ |^3\!D_2\rangle \end{bmatrix}.
\end{gather}
Then the mixing angle for $L=2$ can be expressed as $\alpha=\arctan\sqrt{2/3}=39.23\degree$. So in this definition $D_{2}$ corresponds to the $|J^P,j_\ell\rangle=|2^-,{5/2}\rangle$ state and $D'_{2}$ corresponds to  the $|2^-,{3/2}\rangle$ state. In this work the same mixing angle will also be used for $2^-$ states $D^{(\prime)}_{s2}$ and $B^{(\prime)}_{(s)2}$. Here the mixing angle is the ideal case predicted by the HEQT in the limit of $m_Q\To\infty$. The dependence for decay widths varying over the mixing angle can be seen in equations~(\ref{E-mix1}) and (\ref{E-mix2}).

The wave functions of $^1\!D_2$ and $^3\!D_2$ states can be achieved by solving the corresponding Salpeter equations directly. Then the amplitude for physical $2^-$ states can be considered as the mixing of the transition amplitudes for $^1\!D_2$ and $^3\!D_2$ states, namely
\begin{equation}\label{mix-d2}
\begin{aligned}
\mathcal{M}^\mu(D_2)  = +\cos\alpha~\mathcal{M}^{\mu}(^1\!D_2) + \sin \alpha~\mathcal{M}^{\mu}(^3\!D_2),\\
\mathcal{M}^\mu(D'_2) = -\sin\alpha~\mathcal{M}^{\mu}(^1\!D_2) + \cos \alpha~\mathcal{M}^{\mu}(^3\!D_2).
\end{aligned}
\end{equation}

By using Eq.~(\ref{mix-d2}), replacing the final state's wave function $\psi_D(^3\!D_3)$ by $\psi_D(^1\!D_2)$ and $\psi_D(^3\!D_2)$, and then repeating the above procedures for $^3\!D_3$ state, we can get the form factors $s_i$ for $D_2$ and $t_i$ for $D'_2$ defined in Eq.~(\ref{form}). 

\subsection{Form factors}   
To solve the Salpeter equations, in this work we choose the Cornell potential as the inner interaction kernel as before~\cite{PLB584-2004}, which is a linear scalar potential plus a vector interaction potential as below 
\begin{equation}\label{Cornell}
\begin{aligned}
V(\bm q)&=(2\uppi)^3V_s(\bm q)+\gamma^0\otimes\gamma_0(2\uppi)^3V_v(\bm q),\\
V_s(\bm q)&=-(\frac{\lambda}{\alpha}+V_0)\delta^3(\bm q)+\frac{\lambda}{\uppi^2(\bm q^2+\alpha^2)^2},\\
V_v(\bm q)&=-\frac{2\alpha_s(\bm q)}{3\uppi^2(\bm q^2+\alpha^2)}, ~
\alpha_s(\bm q)=\frac{12\uppi}{27\ln(a+\frac{\bm q^2}{\Lambda^2_\mathrm{QCD}})}.
\end{aligned}
\end{equation}
In above equations, the symbol $\otimes$ denotes that the Salpeter wave function is sandwiched between the two $\gamma^0$ matrices. The model parameters we used are the same with before~\cite{JPG40-2013}, which read
\begin{align*}
  a&=e=2.7183,    	 &\alpha&=0.060~\si{GeV}, &\lambda&=0.210~\si{GeV}^2, \\
m_u&=0.305~\si{GeV},  &m_d&=0.311~\si{GeV},    & m_s&=0.500~\si{GeV},\\
m_c&=1.62~\si{GeV},   &m_b&=4.96~\si{GeV},    &\Lambda_\text{QCD}&=0.270~\si{GeV}.
\end{align*}
The free parameter $V_0$ is fixed by fitting the mass eigenvalue to experimental value. 

With the numerical Salpeter wave function we can obtain the form factors.

\begin{figurehere}
\centering
\subfigure[Form Factors for $\bar B\To D_{2}(2^-)$.]{\includegraphics[width=0.48\textwidth]{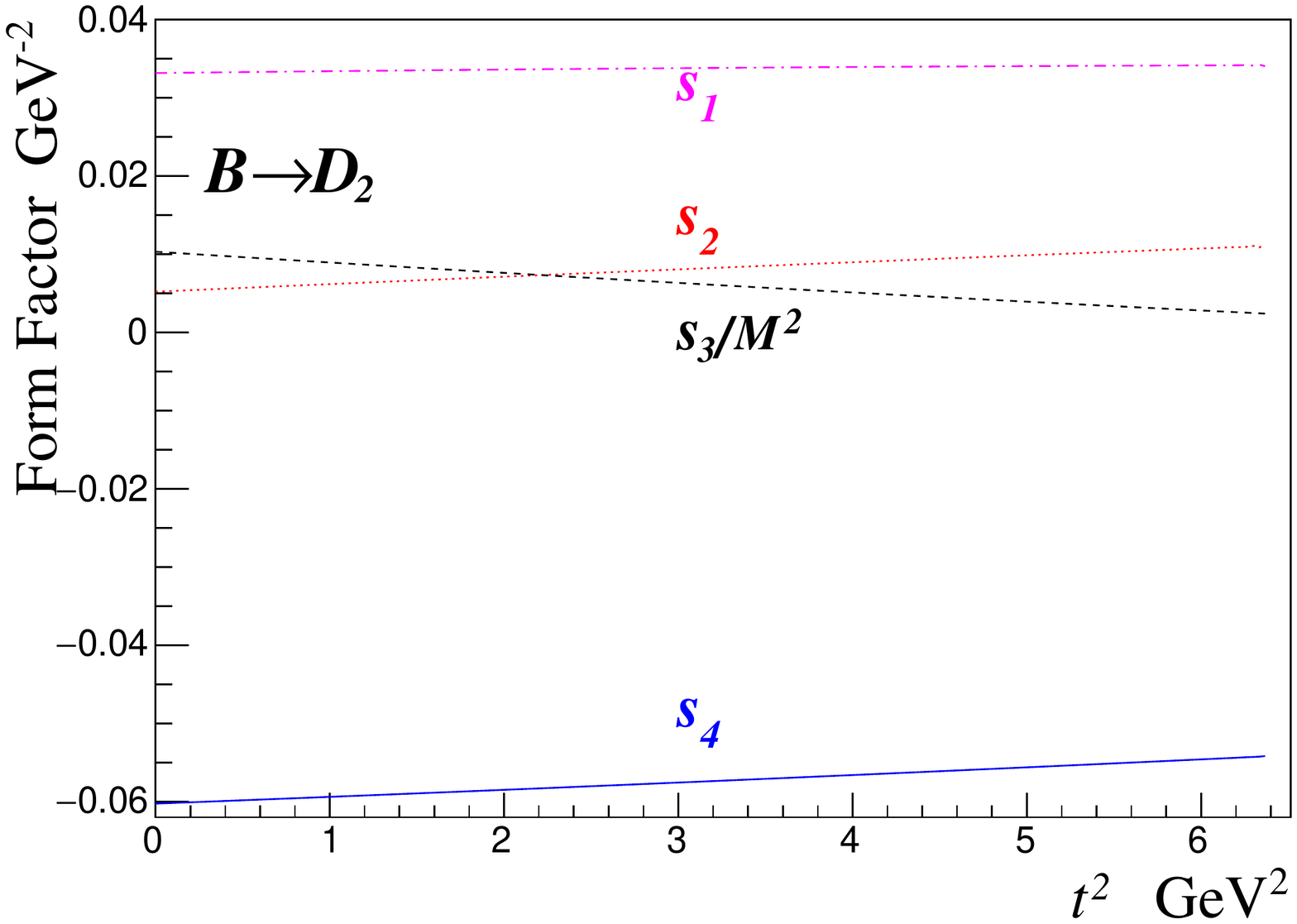} \label{Fig-D2a}}
\subfigure[Form Factors for $\bar B\To D'_{2}(2^-)$.]{\includegraphics[width=0.48\textwidth]{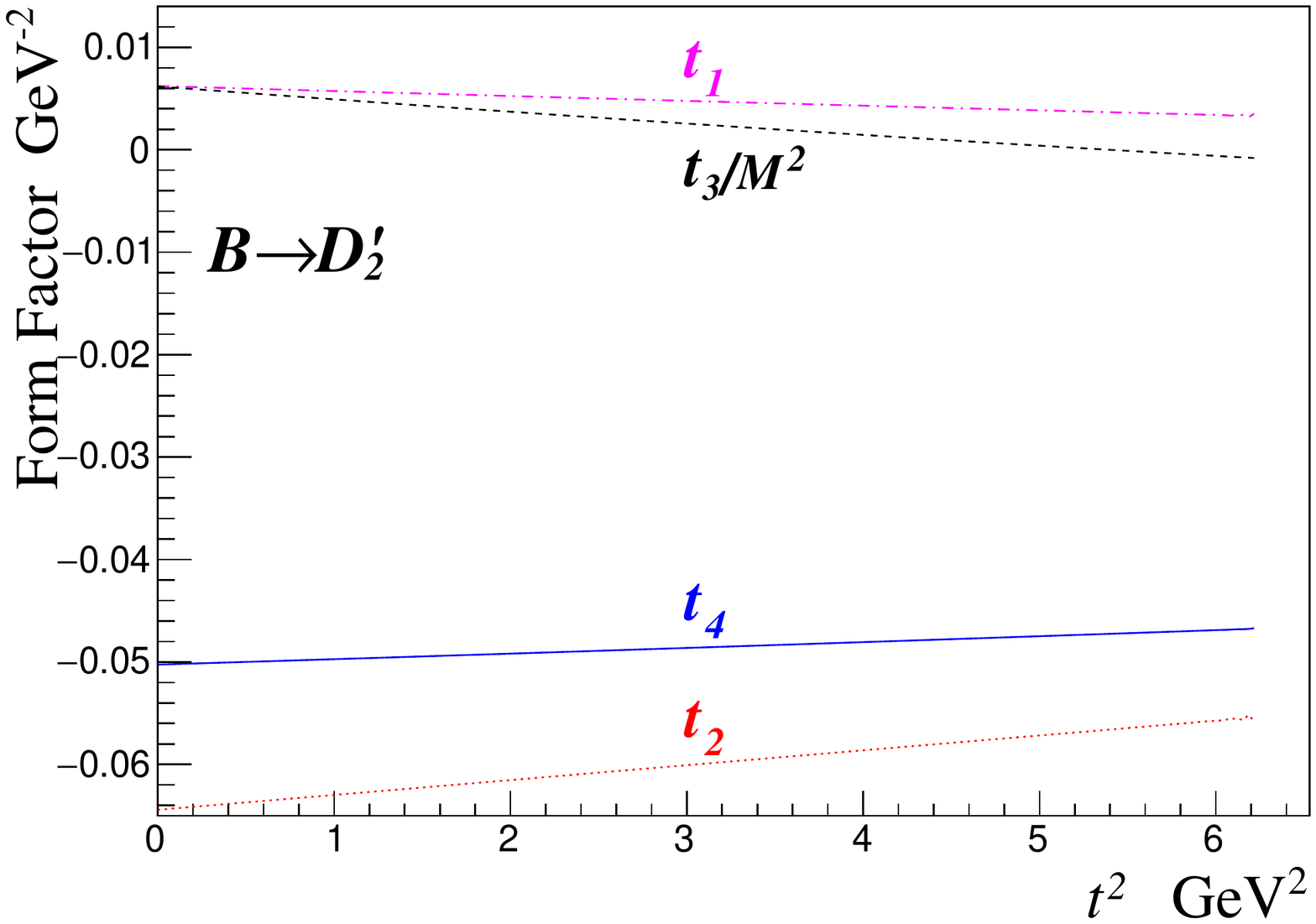} \label{Fig-D2b}}\\
\subfigure[Form Factors for $\bar B\To D^*_{3}(3^-)$.]{\includegraphics[width=0.48\textwidth]{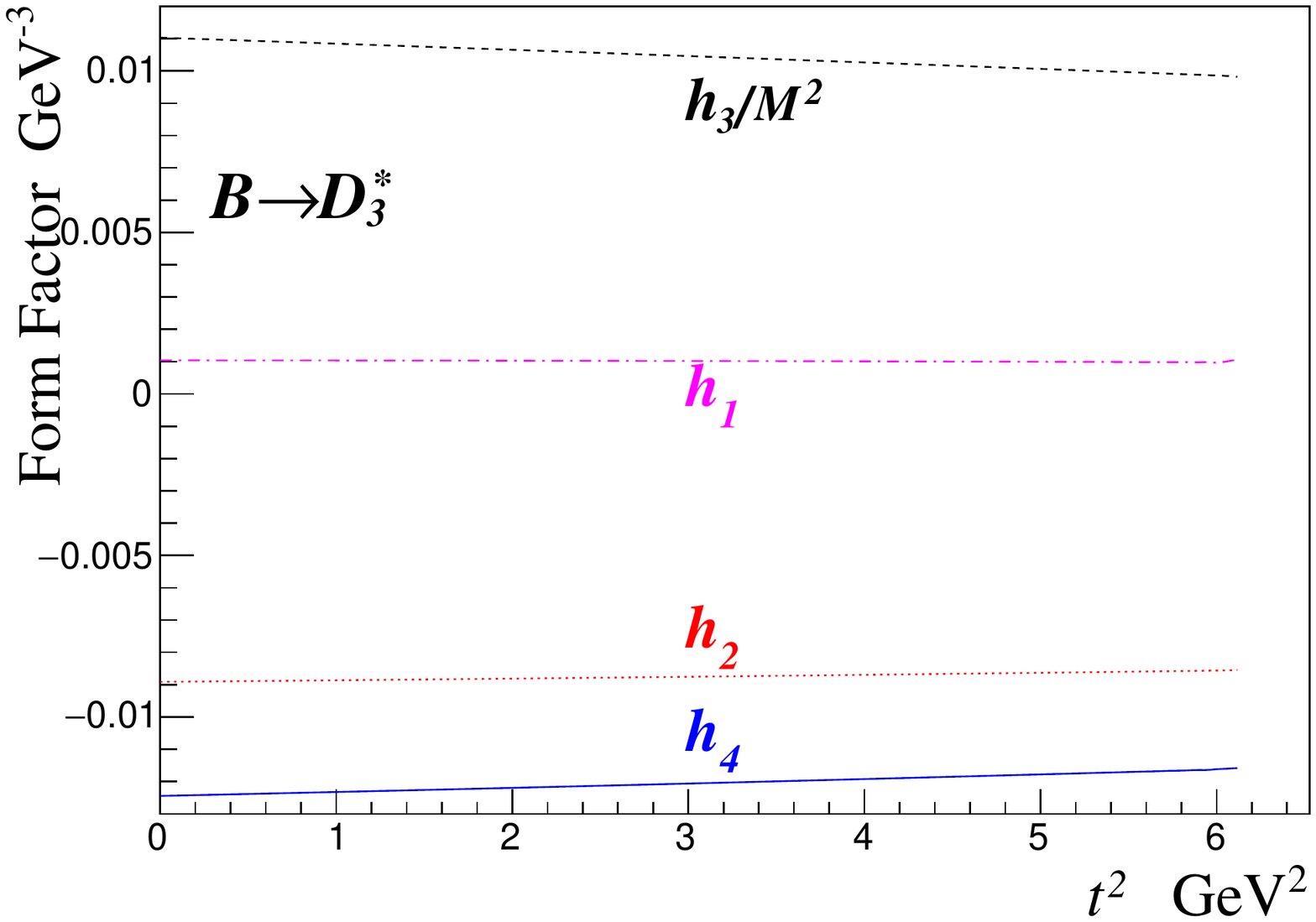} \label{Fig-D3}}
\subfigure[Form Factors for $B^-_c\To \bar D_{2}(2^-)$.]{\includegraphics[width=0.48\textwidth]{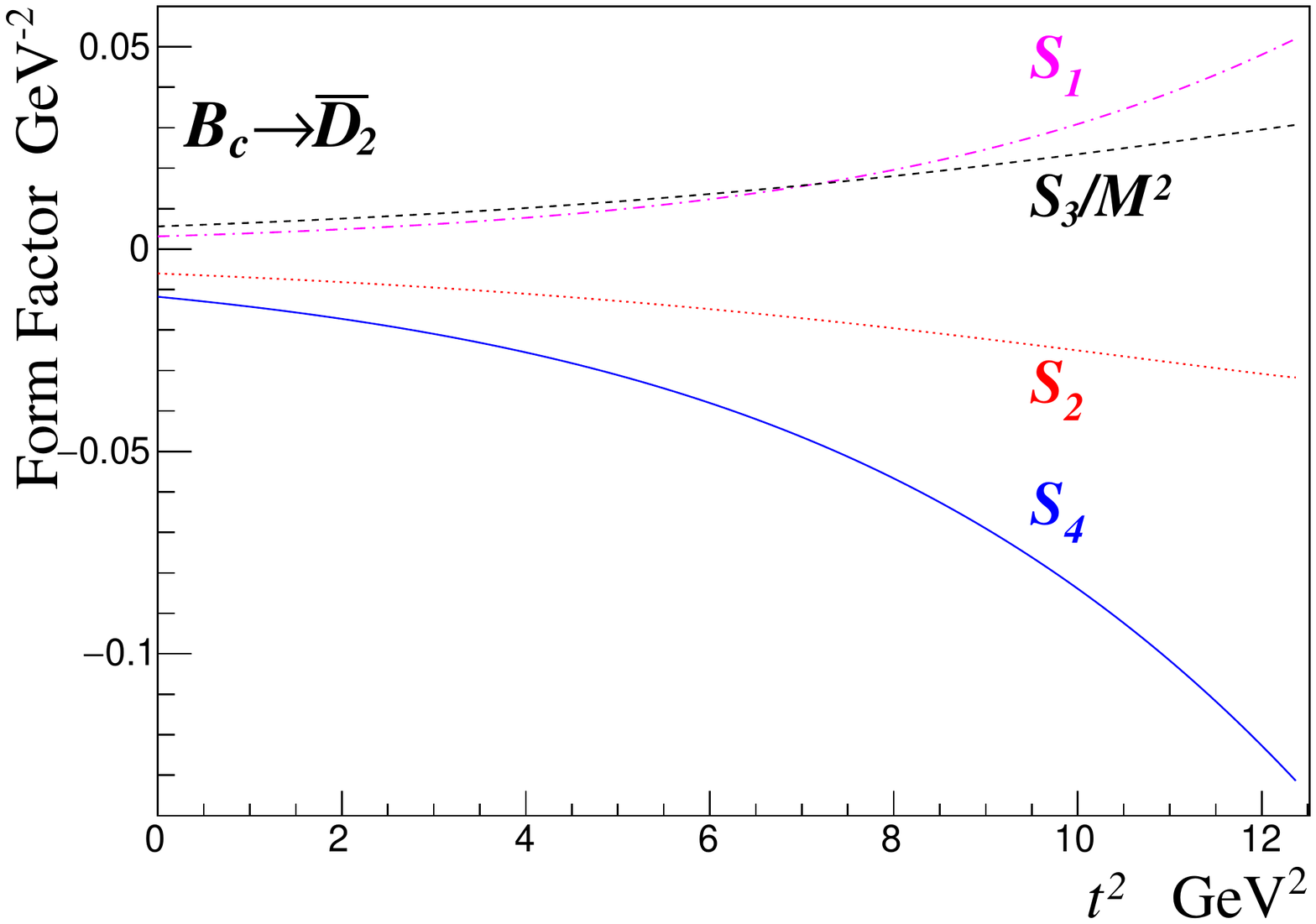} \label{Fig-D2a-Bc}}\\
\subfigure[Form Factors for $B^-_c\To \bar D'_{2}(2^-)$.]{\includegraphics[width=0.48\textwidth]{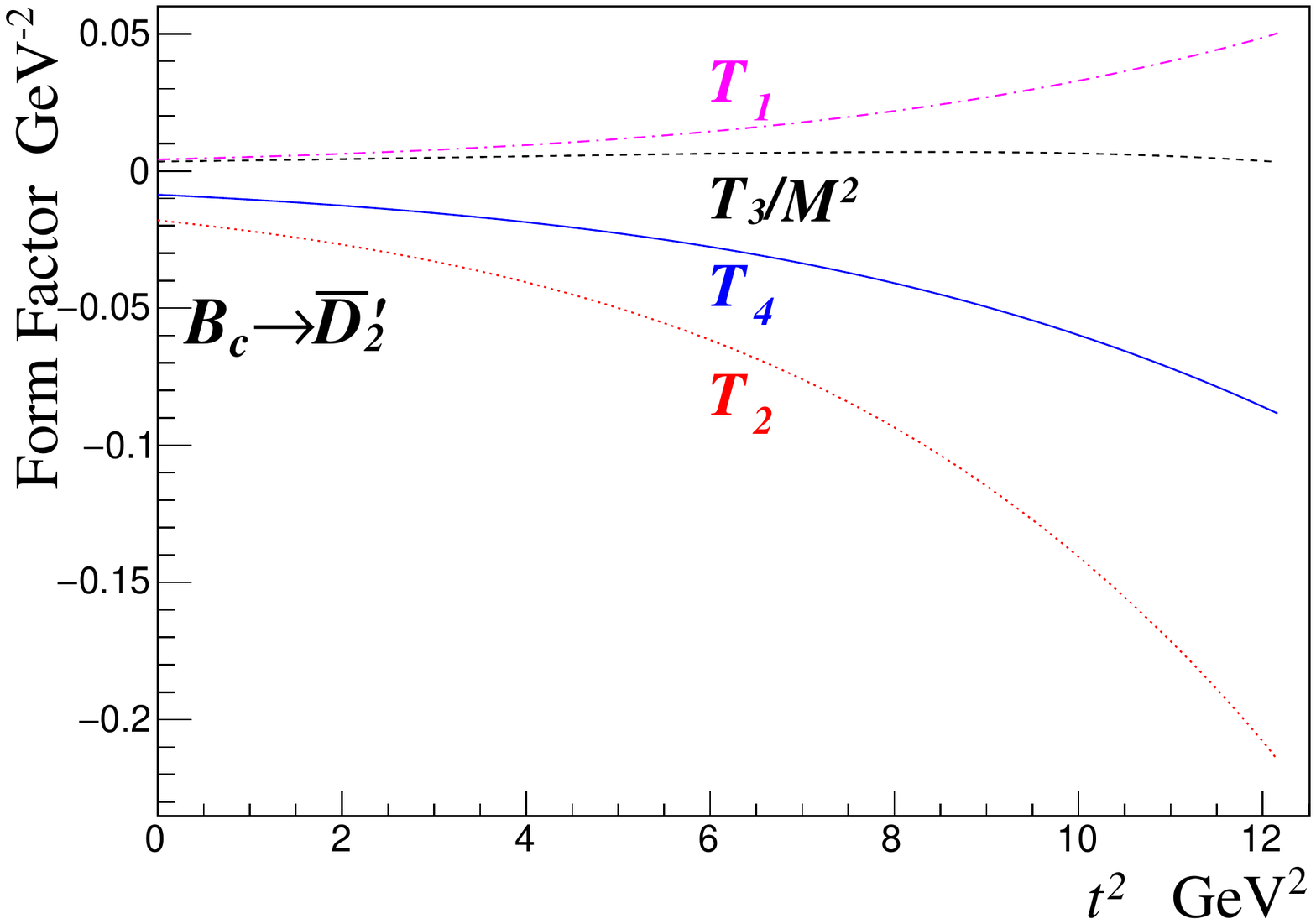} \label{Fig-D2b-Bc}}
\subfigure[Form Factors for $B^-_c\To \bar D^*_{3}(3^-)$.]{\includegraphics[width=0.48\textwidth]{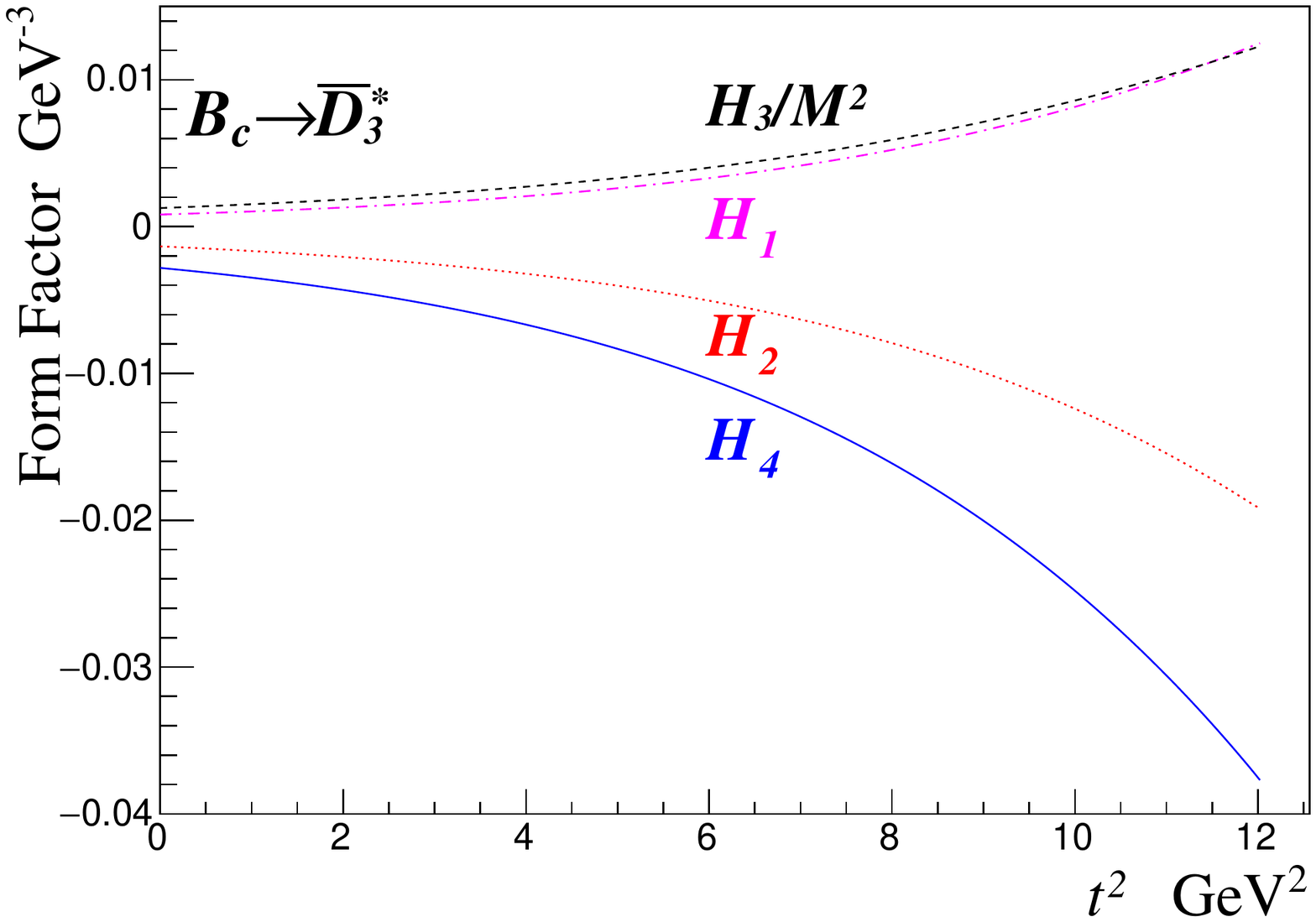} \label{Fig-D3-Bc}}
\caption{Form factors for transitions $\bar{B}\To D^{(*)}_J~(2^-, 3^-)$ and $B^-_c\To \bar D^{(*)}_J~(2^-, 3^-)$. $t^2=(P-P_F)^2$ denotes the square of momentum transfer. To make the dimension consistent, $s_3$, $t_3$ and $h_3$ are divided by $M_{\bar B}^2$, $S_3$, $T_3$ and $H_3$ are divided by  $M_{B_c}^2$.}\label{Fig-form}
\end{figurehere}

Here we plot the $\bar B\To D^{(*)}_J$ form factors $s_i$, $t_i$, and $h_i~(i=1,2,3,4)$ changing with the square of momentum transfer $t^2=(P-P_F)^2$ in \autoref{Fig-D2a}~$\sim$~\autoref{Fig-D3}, respectively, where $s_3$, $t_3$, and $h_3$ are divided by $M^2_{\bar B}$ in order to keep the dimension consistent. \autoref{Fig-D2a-Bc}~$\sim$~\autoref{Fig-D3-Bc} are the distribution of form factors $S_i$, $T_i$, and $H_i$ for $B^-_c\To \bar D^{(*)}_J$ transitions. Also we divided $S_3$, $T_3$, and $H_3$ by $M^2_{B_c}$ to keep the dimension consistent.
From~\autoref{Fig-form}, we can see that in all the range concerned the form factors are quite smooth along with $t^2$. And for transitions $\bar B\To D^{(*)}_J$, the form factors change slowly and almost linearly when $t^2$ varies from $0$ to $(M-M_F)^2$. For transitions $B^-_c\To \bar D^{(*)}_J$, the form factors change dramatically over $t^2$, especially in the range with large momentum transfer.

\section{Numerical Results and Discussions}\label{Sec-4}
Firstly we specify the meson mass, lifetime, CKM matrix elements and decay constants used in this work. For the mass of $\bar{B}$, $\bar B_s$, and $B_c$ mesons we take the values from PDG~\cite{PDG-2014}. We follow the mass predictions and $J^P$ assignments of Ref.~\cite{PRD93-2016} for $D$-wave charm and charm-strange mesons. For $D$-wave bottom mesons $B_2$, $B'_2$, and $B^*_3$ we use the average values of Ref.~\cite{Dev-2015} and Ref.~\cite{Fau-2010}. Predictions of Ref.~\cite{Fau-2010} and Ref.~\cite{Dev-2012} are averaged to achieve the mass of $D$-wave bottom-strange mesons $B_{s2}$, $B'_{s2}$, and $B^*_{s3}$. These mass values we used can been seen below
\begin{alignat*}{8}
&M_{B}  	&=5.280~\si{GeV}, \quad &M_{B_s} 	&=5.367~\si{GeV}, \quad &M_{B_c}  	&=6.276~\si{GeV},\\
&M_{D_2}	&=2.750~\si{GeV}, \quad &M_{D'_2}   	&=2.780~\si{GeV}, \quad &M_{D^*_3}	&=2.800~\si{GeV},\\
&M_{D_{s2}}&=2.846~\si{GeV}, \quad &M_{D'_{s2}}	&=2.872~\si{GeV}, \quad &M_{D^*_{s3}}	&=2.860~\si{GeV},\\
&M_{B_{2}}	&=6.060~\si{GeV}, \quad &M_{B'_{2}}	&=6.100~\si{GeV}, \quad &M_{B^*_{3}}	&=6.050~\si{GeV},\\
&M_{B_{s2}}&=6.150~\si{GeV}, \quad &M_{B'_{s2}}	&=6.210~\si{GeV}, \quad &M_{B^*_{s3}}	&=6.190~\si{GeV}.
\end{alignat*}
The lifetime of initial mesons we used are as below~\cite{PDG-2014}
\begin{gather*}
\tau_{\bar B}=1.519\times 10^{-12}~\si{s},  \quad \tau_{\bar B_s}=1.512\times 10^{-12}~\si{s}, \quad \tau_{ B_c}=0.452\times 10^{-12}~\si{s}.
\end {gather*}
The involved CKM matrix element values are~\cite{PDG-2014}
\begin{gather*}
|V_{ud}|=0.974,~ |V_{us}|=0.225, ~ |V_{ub}|=0.0042, ~ |V_{cd}|=0.23, ~|V_{cs}|=1.006, ~|V_{cb}|=0.041.
\end{gather*}
In the calculation of non-leptonic decays, the decay constants we used are~\cite{PDG-2014,PRD73-2006}
\begin{gather*}
f_\pi=130.4 ~\si{MeV},  ~f_K=156.2 ~\si{MeV},~f_\rho=210 ~\si{MeV},~f_{K^*}=217~ \si{MeV}.
\end{gather*}

For the theoretical uncertainties, here we will just discuss the dependence of the final results on our model parameters $\lambda,~\Lambda_\text{QCD}$ in the Cornell potential, and the constituent quark mass $m_b,~m_c,~m_s,~m_d$ and $m_u$. The theoretical errors, induced by these model parameters, are determined by varying every parameter by $\pm5\%$, and then scanning the parameters space to find the maximum deviation. Generally, this theoretical uncertainties can amount to $10\%\sim30\%$ for the semi-leptonic decays. The theoretical uncertainties show the robustness of the numerical algorithm.

\subsection{Lepton spectra and $A_{FB}$}
The distribution of $\bar B$ and $B^-_c$ decay width $\Gamma$ varying along with $\cos \theta$ for $e$ and $\tau$ modes can be seen in \autoref{Fig-cos}, from which we can see that, for $\bar B$ decays, the distribution of semi-electronic decay widths are much more symmetric than that for the semi-taunic mode. These asymmetries over $\cos \theta$ can also be reflected by the forward-backward asymmetries $A_{FB}$, which are showed in \autoref{Tab-AFB}. We can see that $A_{FB}$ is sensitive to lepton mass $m_\ell$ and is the monotonic function of $m_\ell$. Considering the absolute values of $A_{FB}$, we find that for $\bar B\To D^{(*)}_J$ and $B^-_c\To \bar D^{(*)}_J$, the $\mu$ decay mode has the smallest $|A_{FB}|$.\\

\begin{figurehere}
\centering
\subfigure[Angular spectrum for $\bar B\To D^{(*)}_Je\bar \nu$.]      {\includegraphics[width=0.45\textwidth]{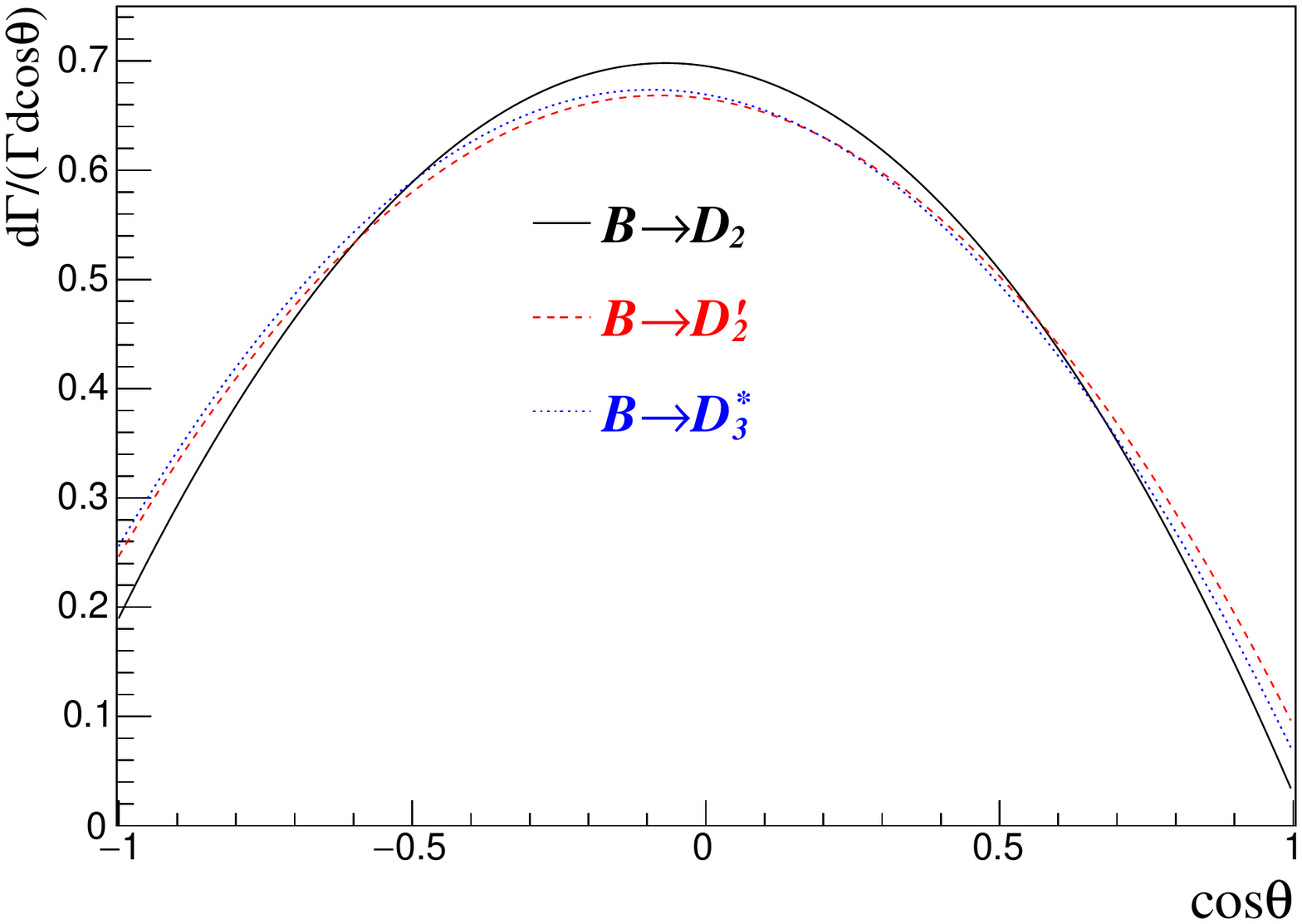} \label{Fig-dwe-cos}}
\subfigure[Angular spectrum for $\bar B\To D^{(*)}_J\tau\bar \nu$.]   {\includegraphics[width=0.45\textwidth]{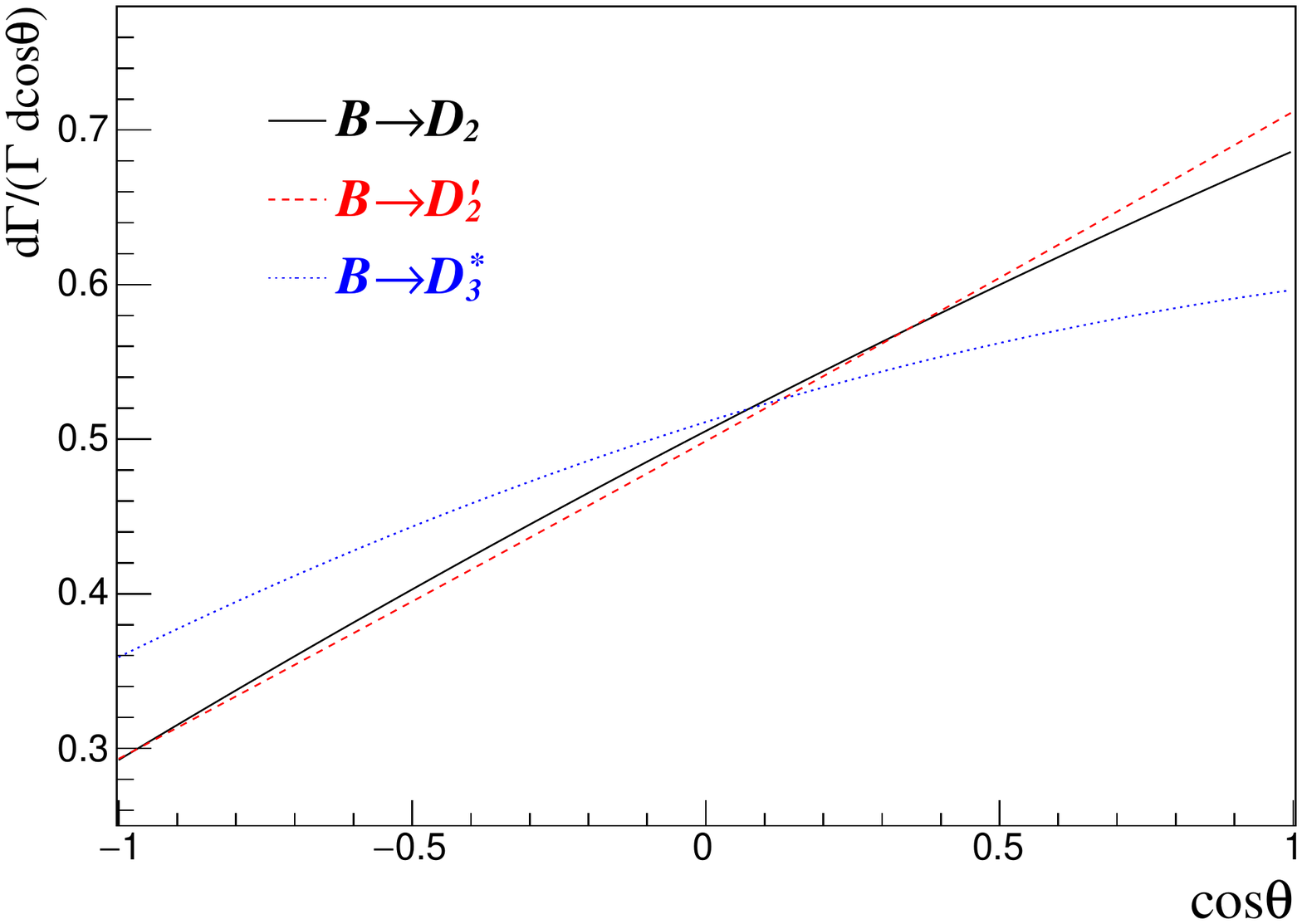} \label{Fig_dwt-cos}}\\
\subfigure[Angular spectrum for $B^-_c\To \bar D^{(*)}_Je\nu$ mode.]      {\includegraphics[width=0.45\textwidth]{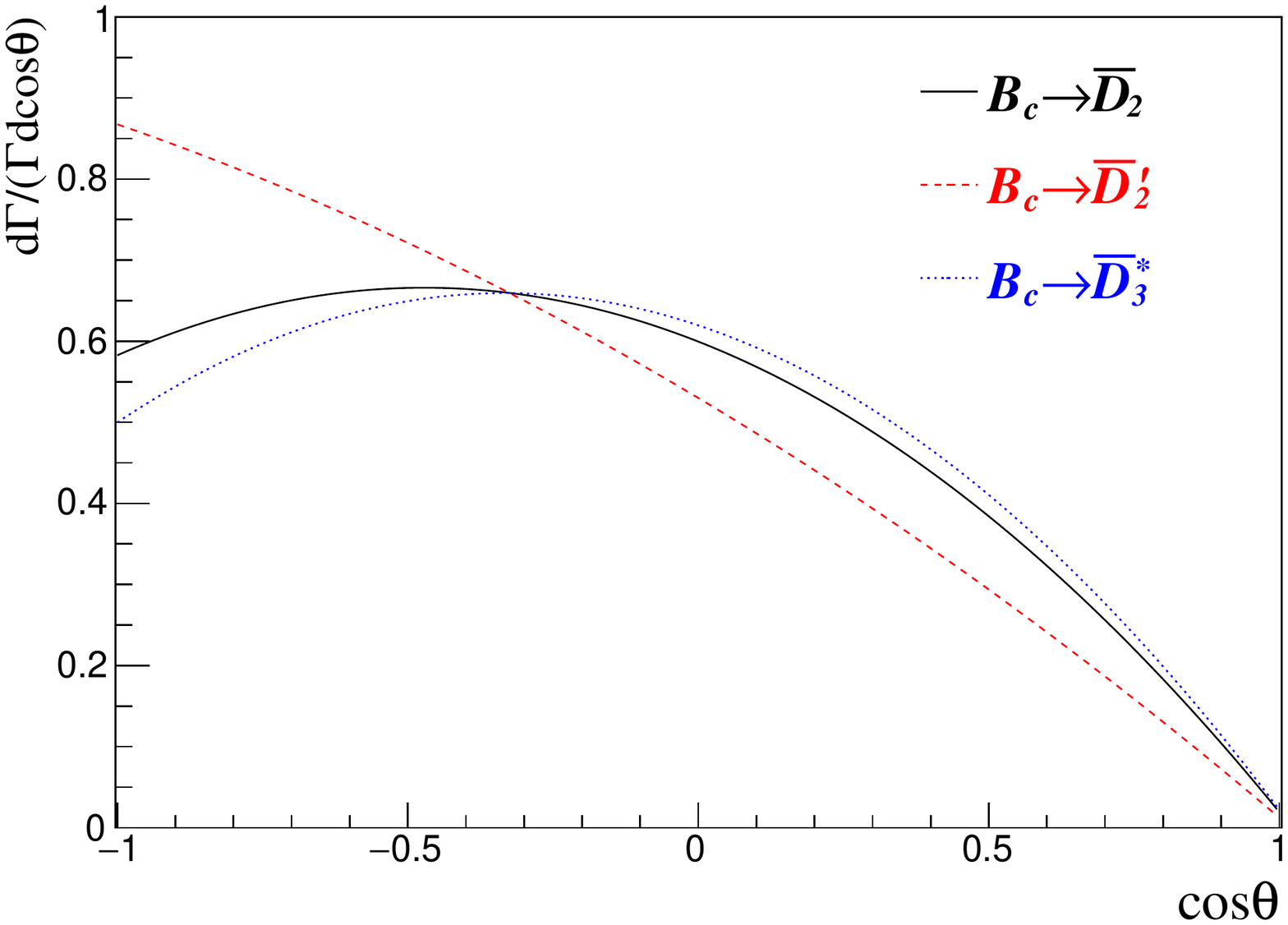} \label{Fig-dwe-cos-Bc}}
\subfigure[Angular spectrum for $B^-_c\To \bar D^{(*)}_J\tau\nu$ mode.]   {\includegraphics[width=0.45\textwidth]{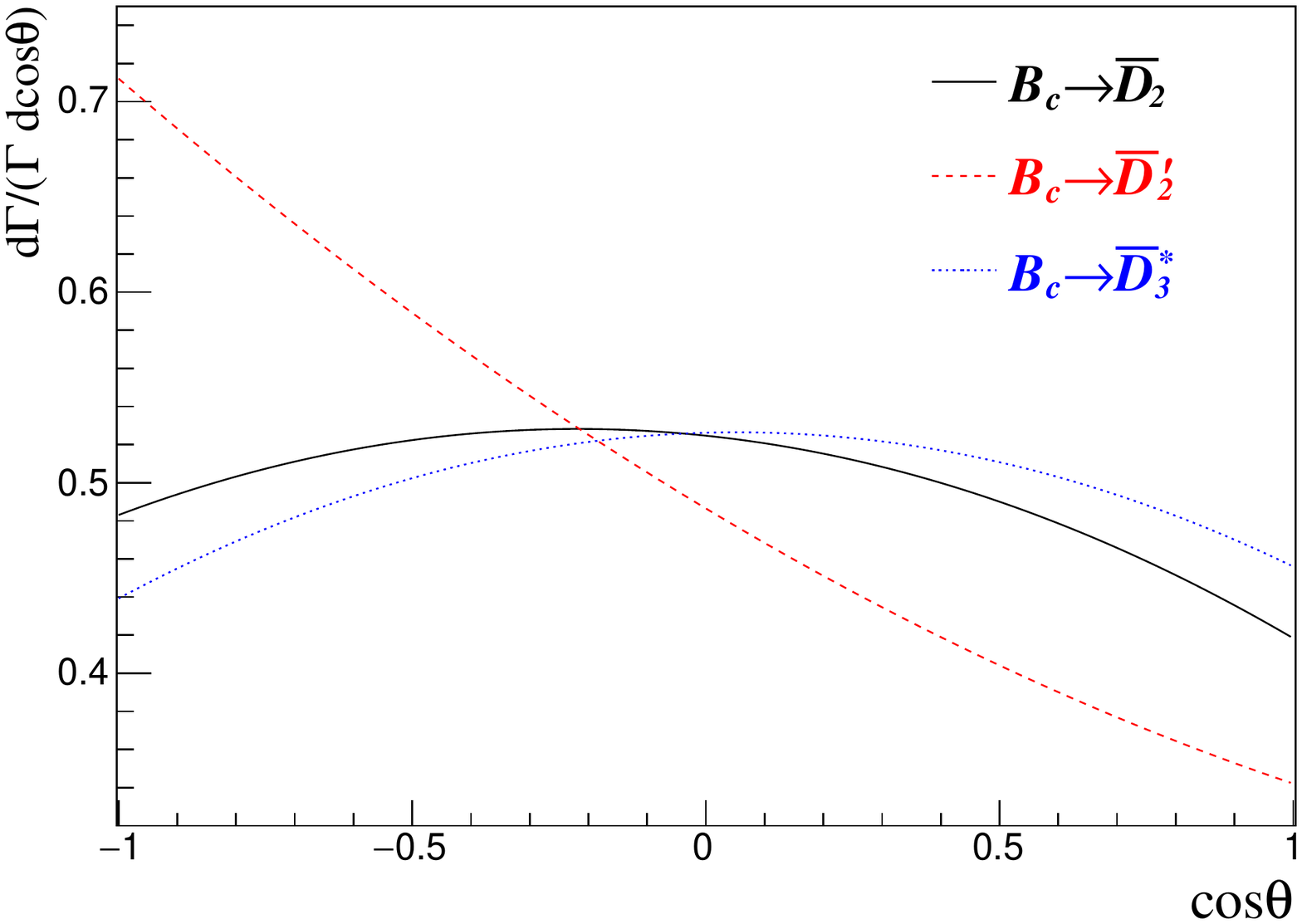} \label{Fig_dwt-cos-Bc}}
\caption{The spectra of relative width vs $\cos\theta$ for semi-leptonic decays $\bar B\To D^{(*)}_J$ and $B^-_c\To \bar D^{(*)}_J$. $\theta$ is the angle between charged lepton $\ell$ and final charmed meson in the rest frame of $\ell\bar{\nu}$ pair.} \label{Fig-cos}
\end{figurehere}

\begin{table}[ht]
\caption{$A_{FB}$ for semi-leptonic decays of $\bar B, \bar B_s$ and $B_c$ to $D$-wave heavy-light mesons.}\label{Tab-AFB}
\vspace{0.2em}\centering
\begin{tabular}{crcrcr}
\toprule[2pt]
       Channels              	  &  $A_{FB}$   & Channels						& $A_{FB}$ 	& Channels						&  $A_{FB}$ \\
\midrule[1.5pt]
$\bar B\To \bar D_2e\bar \nu$	  & $ -0.08$	& $\bar B\To \bar D'_2e\bar \nu$	& $ -0.08$		& $\bar B\To \bar D^*_3e\bar \nu$ 	& $-0.10$\\
$\bar B\To \bar D_{2}\mu\bar \nu$ & $ -0.05$ 	& $\bar B\To \bar D'_{2}\mu\bar \nu$	& $ -0.05$      & $\bar B\To \bar D^*_{3}\mu\bar \nu$	& $-0.07$ \\
$\bar B\To \bar D_2\tau\bar \nu$  & $  0.20$	& $\bar B\To \bar D'_2\tau\bar \nu$	& $  0.21$		& $\bar B\To \bar D^*_3\tau\bar \nu$	& $0.12$\\
\midrule[1.2pt]
$\bar B_s\To D_{s2}e\bar \nu$	  & $ -0.10$	& $\bar B_s\To D'_{s2}e\bar \nu$	& $-0.09$		& $\bar B_s\To D^*_{s3}e\bar \nu$ 	& $ -0.10$\\
$\bar B_s\To D_{s2}\mu\bar \nu$	  & $ -0.07$	& $\bar B_s\To D'_{s2}\mu\bar \nu$	& $-0.06$		& $\bar B_s\To D^*_{s3}\mu\bar \nu$ 	& $ -0.08$\\
$\bar B_s\To D_{s2}\tau\bar \nu$  & $  0.17$	& $\bar B_s\To D'_{s2}\tau\bar \nu$	& $ 0.20$		& $\bar B_s\To D^*_{s3}\tau\bar \nu$ 	& $ 0.11$\\
\midrule[1.2pt]
$B_c^-\To \bar D_2e\bar \nu$	  & $ -0.28$	& $B_c^-\To \bar D'_2e\bar \nu$	  	& $-0.43$		& $B_c^-\To \bar D^*_3e\bar \nu$ 	& $ -0.24$\\
$B_c^-\To \bar D_{2}\mu\bar \nu$  & $ -0.28$ 	& $B_c^-\To \bar D'_{2}\mu\bar \nu$	& $-0.42$  	& $B_c^-\To \bar D^*_{3}\mu\bar \nu$	& $ -0.23$ \\
$B_c^-\To \bar D_2\tau\bar \nu$   & $ -0.03$	& $B_c^-\To \bar D'_2\tau\bar \nu$	& $-0.19$		& $B_c^-\To \bar D^*_3\tau\bar \nu$	& $ -0.01$\\
\midrule[1.2pt]
$B_c^+\To  B_2e^+\nu$	  	  	  & $0.04$  	& $B_c^+\To  B'_2e^+\nu$	  		& $-0.07$ 		& $B_c^+\To  B^*_3e^+\nu$			& $0.03$\\
$B_c^+\To  B_2\mu^+\nu$	   	  & $0.23$  	& $B_c^+\To  B'_2\mu^+\nu$	  		& $0.18$		& $B_c^+\To  B^*_3\mu^+\nu$		& $0.24$\\
\midrule[1.2pt]
$B_c^+\To  B_{s2}e^+\nu$		  & $0.03$  	& $B_c^+\To  B'_{s2}e^+\nu$		& $-0.03$ 		& $B_c^+\To  B^*_{s3}e^+\nu$		& $0.01$\\
\bottomrule[2pt]
\end{tabular}
\end{table}

The spectra of decay widths for $\bar B$ and $B^-_c$ varying along with $|\bm{p}_\ell|$, the absolute value of the three-momentum for charged leptons, are showed in~\autoref{Fig-dwp}. This distribution is almost the same for $\bar B$ decays into $D_2$, $D'_2$ or $D^*_3$. For $B^-_c\To \bar D^{(*)}_J$, the momentum spectrum of $\bar D'_2$ is sharper than that of $\bar D_2$ and $\bar D^*_3$.

\vspace{0.5em}
\begin{figurehere}
\centering
\subfigure[Momentum spectrum for decay $\bar B\To D^{(*)}_Je\bar\nu$.]      {\includegraphics[width=0.48\textwidth]{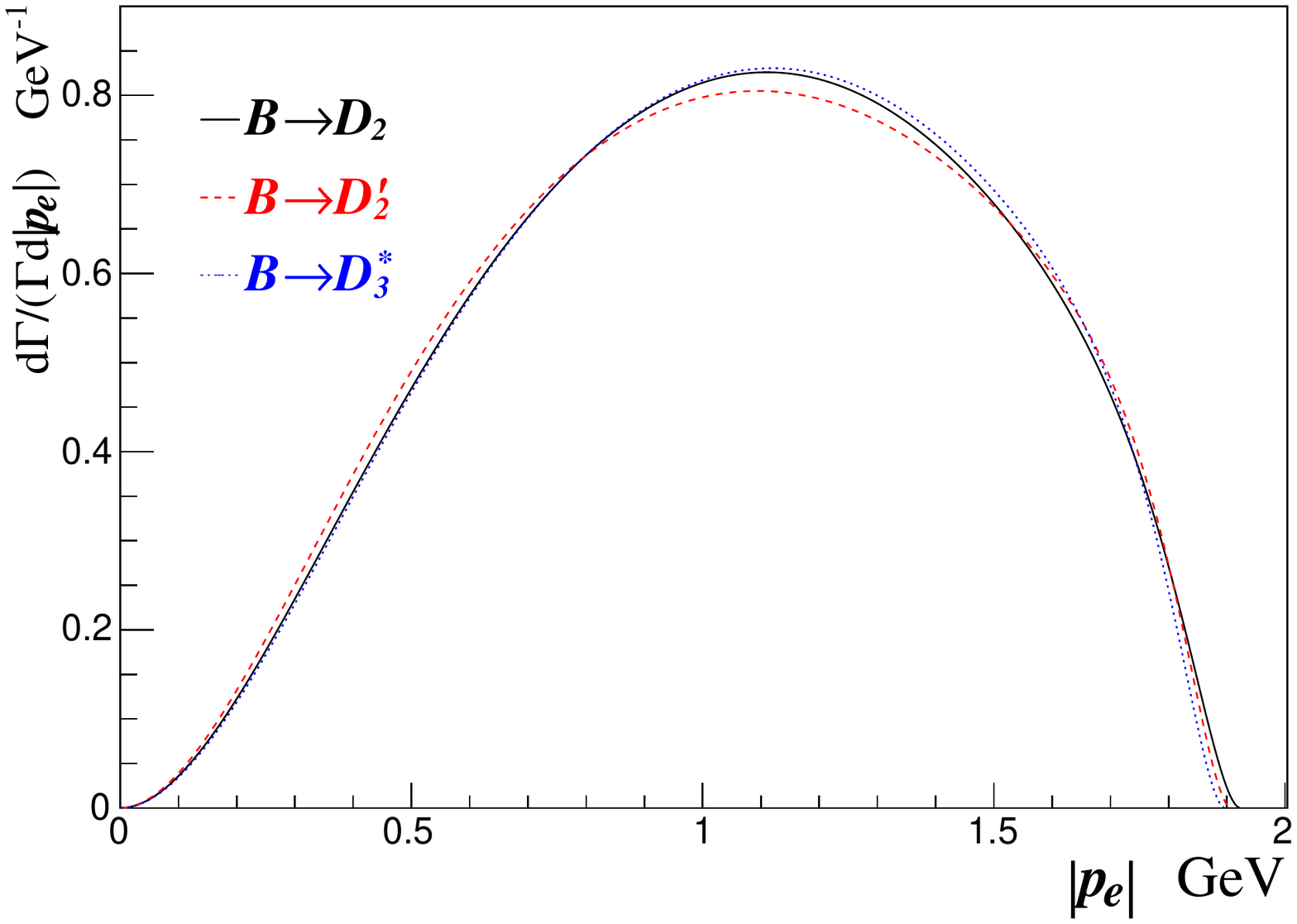} \label{Fig-dwe}}
\subfigure[Momentum spectrum for decay $\bar B\To D^{(*)}_J\tau\bar\nu$.]   {\includegraphics[width=0.48\textwidth]{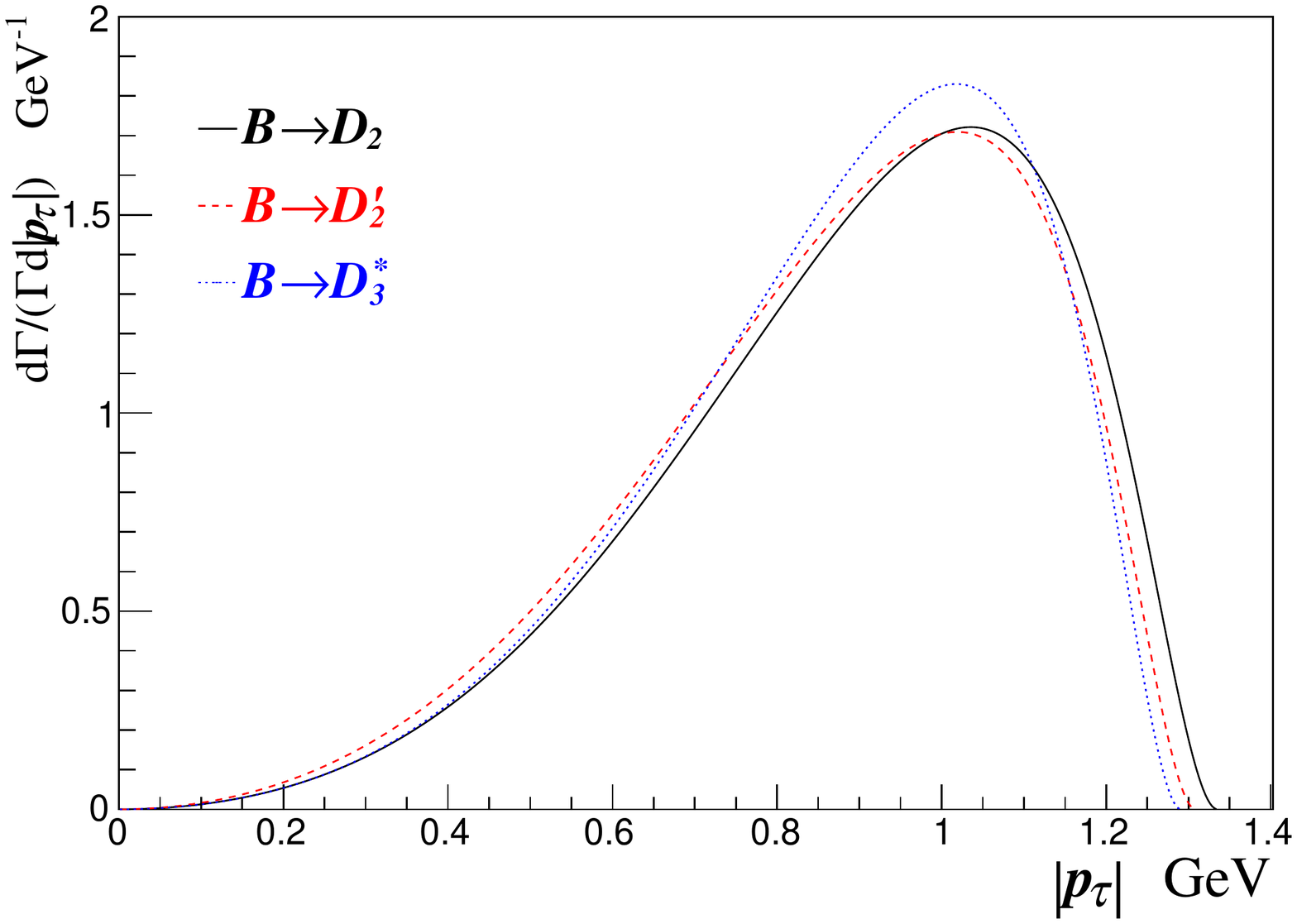} \label{Fig-dwt}}\\
\subfigure[Momentum spectrum for decay $B^-_c\To \bar D^{(*)}_Je\bar\nu$.]      {\includegraphics[width=0.48\textwidth]{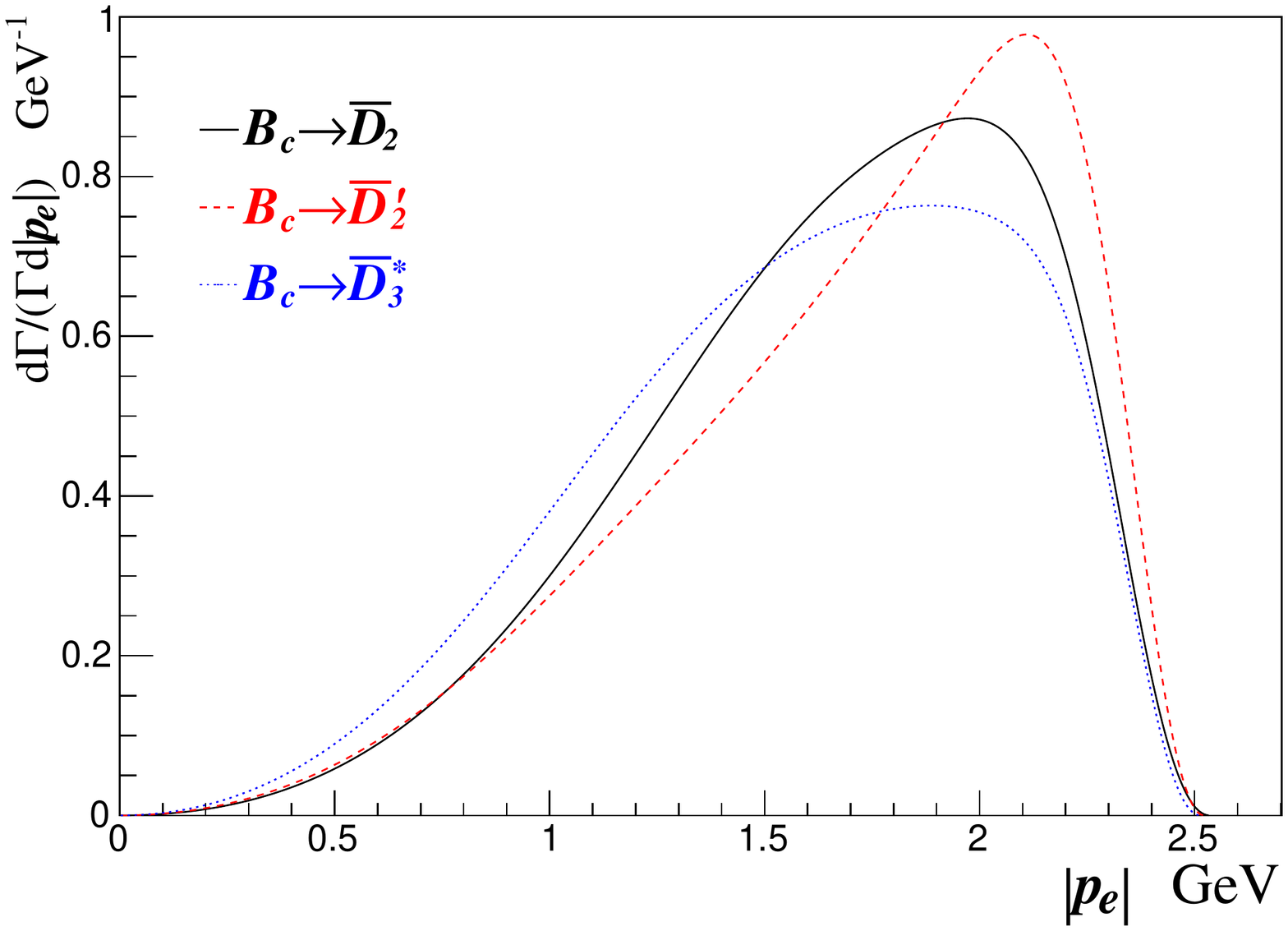} \label{Fig-dwe-Bc}}
\subfigure[Momentum spectrum for decay $B^-_c\To \bar D^{(*)}_J\tau\bar\nu$.]   {\includegraphics[width=0.48\textwidth]{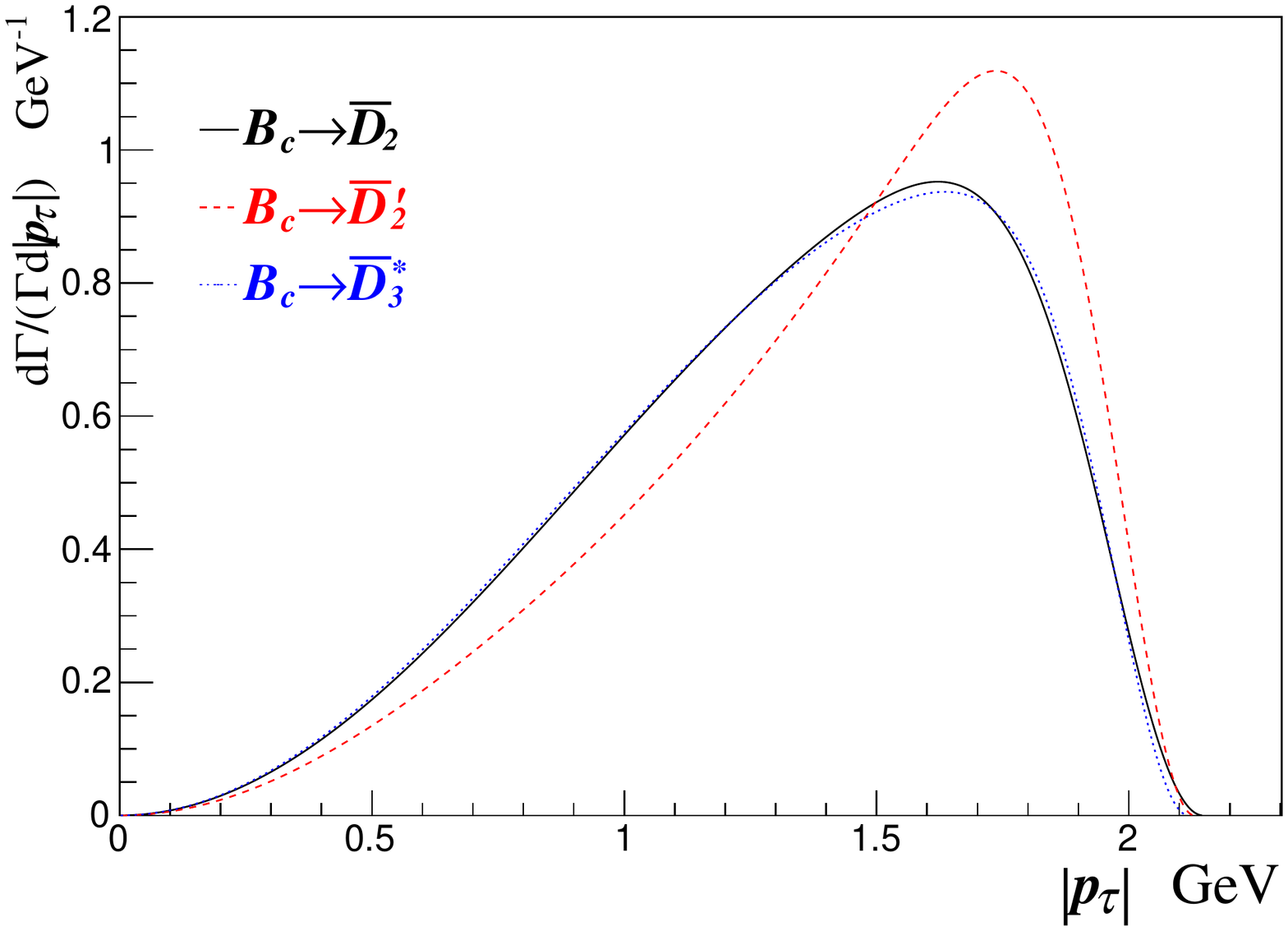} \label{Fig-dwt-Bc}}\\
\caption{The spectra of relative width vs $|\bm{p}_\ell|$, the absolute value of charged lepton's 3-momentum, in transitions $\bar B\To D^{(*)}_J$ and $B^-_c\To\bar D^{(*)}_J$. }\label{Fig-dwp}
\end{figurehere}
\vspace{0.5em}

\subsection{Branching ratios of semi-leptonic decays}
The semi-electronic decay widths we got are $\Gamma(\bar B\To D_2e\bar\nu_e)=4.9\times10^{-16}$~\si{GeV}, $\Gamma(\bar B\To D'_2e\bar\nu_e)=1.8\times10^{-16}$~\si{GeV}, and $\Gamma(\bar B\To D^*_3e\bar\nu_e)=4.5\times10^{-16}$~\si{GeV}.
The branching ratios of $\bar B$ to $D$-wave charmed mesons are listed in~\autoref{br-semi-bd}.
We have listed others' results for comparison if available. Our results are about 5 times greater than that in Ref.~\cite{PRD79-2009}.
It's noticeable that our results for decays into $D_2$ and $D'_2$ are in the same order, while in the results of QCD sum rules~\cite{PRD79-2009} $\mathcal{B}(\bar B\To D_2)$ is about 25 times larger than  $\mathcal{B}(\bar B\To D'_2)$. The branching ratios for semi-leptonic decays of $\bar B_s$ into $D_{s2}$, $D'_{s2}$ and $D^*_{s3}$ are listed in~\autoref{br-semi-Bs}. Our results for $\bar B_s$ to $D$-wave charm-strange mesons are also much larger than the results of QCD sum rules in Ref.~\cite{Gan-2015}.

The branching ratios for $B_c$ to $D$-wave $\bar D^{(*)}_J$ are listed in~\autoref{br-semi-Bc}. The branching ratios for semi-leptonic decays of $B^-_c$ to $\bar D_2$ and $\bar D^*_3$ are in the order of $10^{-5}$, and for $B^-_c\To D'_2$ the results are in the order of $10^{-6}$. These results are about 100 times smaller than that for $\bar B_{(s)}$ decays owning to the the different CKM matrix elements. 

For completeness of this research, we also give the corresponding results for $B_c$ to the $D$-wave $B^{(*)}_J$ and $B^{(*)}_{sJ}$ in~\autoref{br-semi-Bc}, although their branching ratios are quite small duo to the tiny phase space. For $D$-wave bottom mesons, the semi-taunic mode is not available and for $D$-wave bottom-strange mesons, both the $\mu$ and $\tau$ modes are unavailable since the constraints of phase space. The branching ratios for $B^+_c\To B^{(*)}_J$ are less than $10^{-8}$ and that for $B^+_c\To B^{(*)}_{sJ}$ are less than $10^{-9}$. Based on our results, the possibilities for the $D$-wave bottomed mesons to be detected in $B_c$ decays are quite small by current experiments.

\begin{table}[ht]
\caption{Branching ratios of $\bar{B}$ semi-leptonic decays with $\tau_{\bar B}=1.519\times 10^{-12}$~\si{s}.}\label{br-semi-bd}
\vspace{0.2em}\centering
\begin{tabular}{llcccc}
\toprule[2pt]
       Channels             &Ours      & Ref.~\cite{PRD79-2009} &Ref.~\cite{PLB478-2000}& Ref.~\cite{PTP91-1994} &Ref.~\cite{PRD54-1996}  \\
\midrule[1.5pt]
$\bar{B}\To D_2  e  \bar{\nu} $    & $1.1\er{0.3}{0.3}{-3}$	       &		$1.5\e{-4}$		&           $1\e{-5}$    &      -            &        -\\
$\bar{B}\To D_2 \mu  \bar{\nu}$    & $1.1\er{0.3}{0.3}{-3}$            &		$1.5\e{-5}$		&           $1\e{-5}$    &      -            &        -\\
$\bar{B}\To D_2 \tau \bar{\nu}$    & $8.0\er{2.0}{2.0}{-6}$            &		-				&           -      	    &      -            &        -\\
$\bar{B}\To D'_2e\bar{\nu}    $    & $4.1\er{0.8}{0.9}{-4}$            &	     $6\e{-6}$      		&           -            &      $2\e{-6}$    &        $6\e{-5}$\\
$\bar{B}\To D'_2 \mu \bar{\nu}$    & $4.1\er{0.8}{0.9}{-4}$            &	     $6\e{-6}$			&           -            &      $2\e{-6}$    &        -\\
$\bar{B}\To D'_2\tau \bar{\nu}$    & $2.7\er{0.4}{0.5}{-6}$            &	     -				&           -            &      -            &        -\\
$\bar{B}\To D^*_3 e  \bar{\nu}$    & $1.0\er{0.2}{0.2}{-3}$            &		$2.1\e{-4}$		&           $1\e{-5}$    &      -            &		-\\
$\bar{B}\To D^*_3\mu\bar{\nu} $    & $1.0\er{0.2}{0.2}{-3}$            &		$2.1\e{-4}$		&           $1\e{-5}$    &      -            &		-\\
$\bar{B}\To D^*_3\tau\bar{\nu}$    & $5.4\er{0.9}{1.0}{-6}$            &		-				&           -      	    &      -            &		-\\
\bottomrule[2pt]
\end{tabular}
\end{table}

\begin{table}[ht]
\caption{Branching ratios of $\bar{B}_s$ semi-leptonic decays with $\tau_{\bar B_s}=1.512\times 10^{-12}$~\si{s}.}\label{br-semi-Bs}
\vspace{0.2em}\centering
\begin{tabular}{llcc}
\toprule[2pt]
       Channels                   &  Ours          	    & Ref.~\cite{Gan-2015}   \\
\midrule[1.5pt]
$\bar{B}_s\to D_{s2}e \bar{\nu}    $    & $1.7\er{0.5}{0.5}{-3}$   &		$1.02\e{-4}$	\\
$\bar{B}_s\to D_{s2} \mu  \bar{\nu}$    & $1.7\er{0.5}{0.4}{-3}$   &		$1.02\e{-4}$\\
$\bar{B}_s\to D_{s2} \tau \bar{\nu}$    & $1.3\er{0.4}{0.4}{-5}$   &		-	\\
$\bar{B}_s\to D'_{s2}e\bar{\nu}    $    & $5.2\er{1.5}{1.6}{-4}$   &		$3.4\e{-7}$\\
$\bar{B}_s\to D'_{s2} \mu \bar{\nu}$    & $5.1\er{1.5}{1.6}{-4}$   &	    	$3.4\e{-7}$\\
$\bar{B}_s\to D'_{s2}\tau \bar{\nu}$    & $3.4\er{1.0}{1.1}{-6}$   &	     -	  \\
$\bar{B}_s\to D^*_{s3} e  \bar{\nu}$    & $1.5\er{0.4}{0.4}{-3}$   &		$3.46\e{-4}$	  \\
$\bar{B}_s\to D^*_{s3}\mu\bar{\nu} $    & $1.4\er{0.4}{0.4}{-3}$   &		$3.46\e{-4}$	  \\
$\bar{B}_s\to D^*_{s3}\tau\bar{\nu}$    & $9.4\er{2.5}{2.8}{-6}$   &		-	  \\
\bottomrule[2pt]
\end{tabular}
\end{table}

\begin{table}[ht]
\caption{Semi-leptonic decay branching ratios of $B_c$ to $D$-wave heavy-light mesons with $\tau_{B_c}=0.452\times 10^{-12}$~\si{s}.}\label{br-semi-Bc}
\vspace{0.2em}\centering
\begin{tabular}{lclclc}
\toprule[2pt]
       Channels              	  &  Br         	& Channels					  	&	Br   		& Channels						& Br \\
\midrule[1.5pt]
$B_c^-\To \bar D_2e\bar \nu$	  & $ 2.2\er{0.4}{0.7}{-5}$	& $B_c^-\To \bar D'_2e\bar \nu$	  	& $4.0\er{0.9}{1.6}{-6}$	& $B_c^-\To \bar D^*_3e\bar \nu$ 	& $ 1.2\er{0.2}{0.4}{-5}$\\
$B_c^-\To \bar D_{2}\mu\bar \nu$  & $2.2\er{0.4}{0.7}{-5}$ 	& $B_c^-\To \bar D'_{2}\mu\bar \nu$	& $4.0\er{0.9}{1.6}{-6}$   & $B_c^-\To \bar D^*_{3}\mu\bar \nu$	& $1.2\er{0.2}{0.4}{-5}$ \\
$B_c^-\To \bar D_2\tau\bar \nu$   & $7.7\er{1.5}{2.6}{-6}$		& $B_c^-\To \bar D'_2\tau\bar \nu$	& $1.2\er{0.3}{0.5}{-6}$	& $B_c^-\To \bar D^*_3\tau\bar \nu$	& $3.1\er{0.7}{1.1}{-6}$\\
$B_c^+\To  B_2e^+\nu$	  	  	  & $9.4\er{1.0}{0.7}{-9}$  	& $B_c^+\To  B'_2e^+\nu$	  		& $1.3\er{0.3}{0.3}{-10}$  & $B_c^+\To  B^*_3e^+\nu$			& $1.4\er{0.3}{0.3}{-10}$\\
$B_c^+\To  B_2\mu^+\nu$	   	  & $1.7\er{0.2}{0.1}{-9}$  	& $B_c^+\To  B'_2\mu^+\nu$	  		& $7.6\er{0.7}{0.7}{-12}$  & $B_c^+\To  B^*_3\mu^+\nu$		& $2.0\er{0.4}{0.4}{-11}$\\
$B_c^+\To  B_{s2}e^+\nu$		  & $3.3\er{0.2}{0.2}{-9}$  	& $B_c^+\To  B'_{s2}e^+\nu$		& $3.2\er{0.4}{0.4}{-12}$  & $B_c^+\To  B^*_{s3}e^+\nu$		& $5.6\er{1.7}{1.7}{-13}$\\
\bottomrule[2pt]
\end{tabular}
\end{table}

The ratio $\mathscr{R}[D^{(*)}_J]$, defined as the ratio of semi-taunic branching fraction over semi-electronic branching fraction for decay $\bar B\To D^{(*)}_J$, namely, $\mathscr{R}[D^{(*)}_{J}]=\frac{\mathcal{B}[\bar B\to D^{(*)}_{J}\tau^-\bar \nu_\tau]}{\mathcal{B}[\bar B\to D^{(*)}_{J}e^-\bar \nu_e]}$, may hint the new physics~\cite{BaBar-2012,Belle-2016}. We present these ratios for decays to $D$-wave charmed mesons in~\autoref{Tab-RD}, from which we can see that, $\mathscr{R}[D^{(*)}_{J}]$ for $\bar B$ decays and $\mathscr{R}[D^{(*)}_{sJ}]$ for $\bar B_s$ decays, are almost the same and in the order of $10^{-3}$, while $\mathscr{R}[\bar D^{(*)}_{J}]$ for $B^-_c$ decays are in the order of $10^{-1}$. This big difference is mainly due to the phase space. By simple integral over the phase space, we can find that, the phase space ratio of semi-taunic decay over semi-electronic decay for $B_c^-$ meson is about 30 times larger than that for $\bar B$ or $\bar B_s$ meson.

\begin{table}[h]
\centering
\caption{$\mathscr{R}[D^{(*)}_J]=\frac{\mathscr{B}[\bar B\to D^{(*)}_J\tau\bar \nu_\tau]}{\mathscr{B}[\bar B\to D^{(*)}_Je\bar \nu_e]}$, $\mathscr{R}[D^{(*)}_{sJ}]=\frac{\mathscr{B}[\bar B_s\to D^{(*)}_{sJ}\tau\bar \nu_\tau]}{\mathscr{B}[\bar B_s\to D^{(*)}_{sJ}e\bar \nu_e]}$, and $\mathscr{R}[\bar D^{(*)}_J]=\frac{\mathscr{B}[B^-_c\to \bar D^{(*)}_J\tau \bar \nu_\tau]}{\mathscr{B}[B_c^-\to \bar D^{(*)}_Je \bar \nu_e]}$, ratios of semi-taunic branching ratio to semi-electronic branching ratio for $\bar B$, $\bar B_s$ and $B^-_c$ to $D$-wave charmed mesons.}\label{Tab-RD}
\vspace{0.2em}
\begin{tabular}{cccccccccc}
\toprule[2pt]
Modes				& $D_2$		& $D'_2$		& $D^*_3$		& $D_{s2}$		& $D'_{s2}$	& $D^*_{s3}$  	 & $\bar D_2$	& $\bar D'_2$   & $\bar D^*_3$  \\
\midrule[1.5pt]
$\mathscr R$		& $0.0071$		& $0.0065$		& $0.0052$		& $0.0079$		& $0.0066$		& $0.0064$	  	 & $0.35$		& $0.29$		& $0.25$\\
\bottomrule[2pt]
\end{tabular}
\end{table}

\begin{figure}[ht]
\centering
\subfigure[$\bar B\To D^{(\prime)}_2e\bar \nu$ decay width vs mixing angle.]{\includegraphics[width=0.48\textwidth]{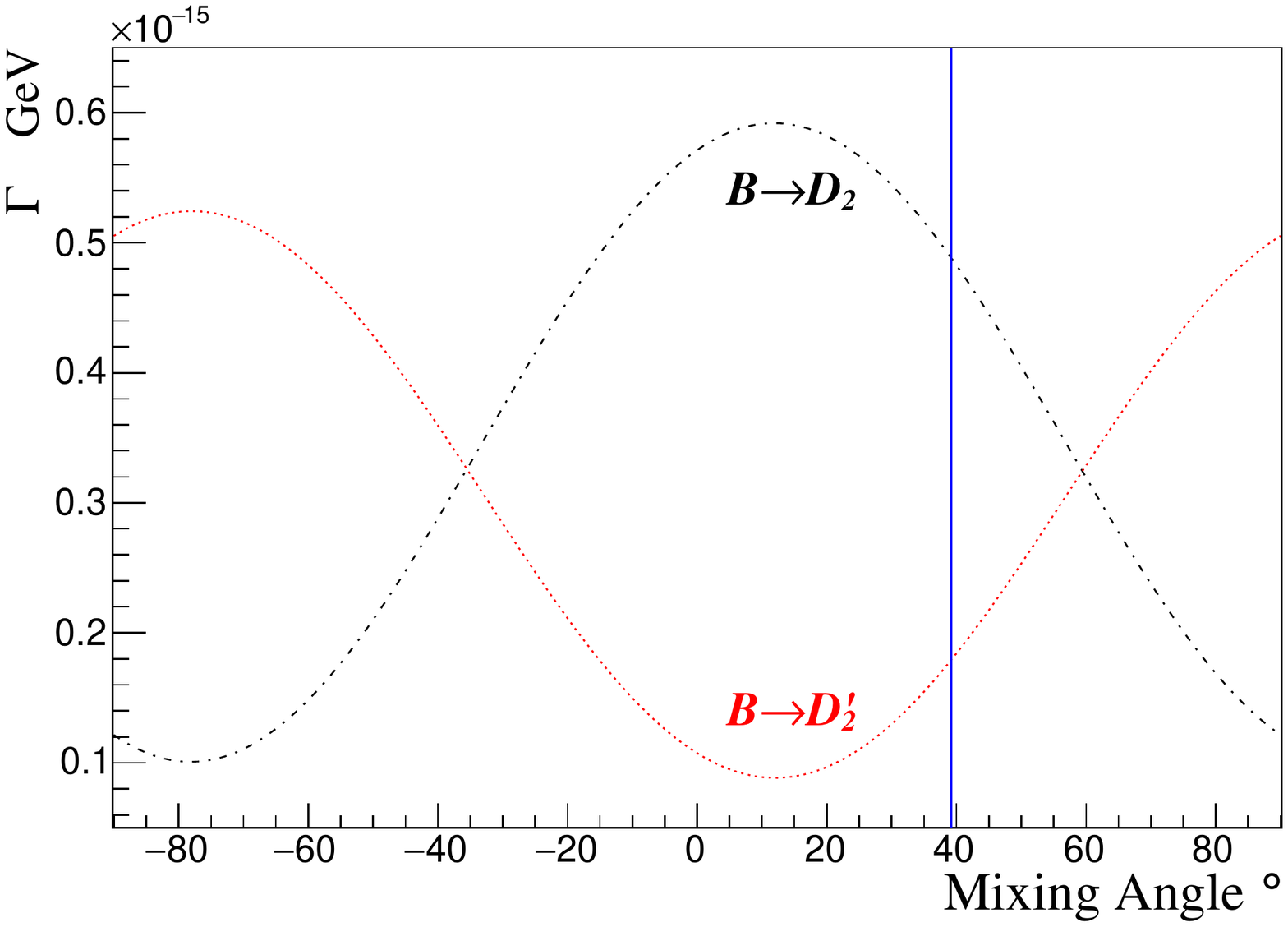} \label{Fig-An1}}
\subfigure[$B^-_c\To \bar D^{(\prime)}_2e\bar \nu$ decay width vs mixing angle.]{\includegraphics[width=0.48\textwidth]{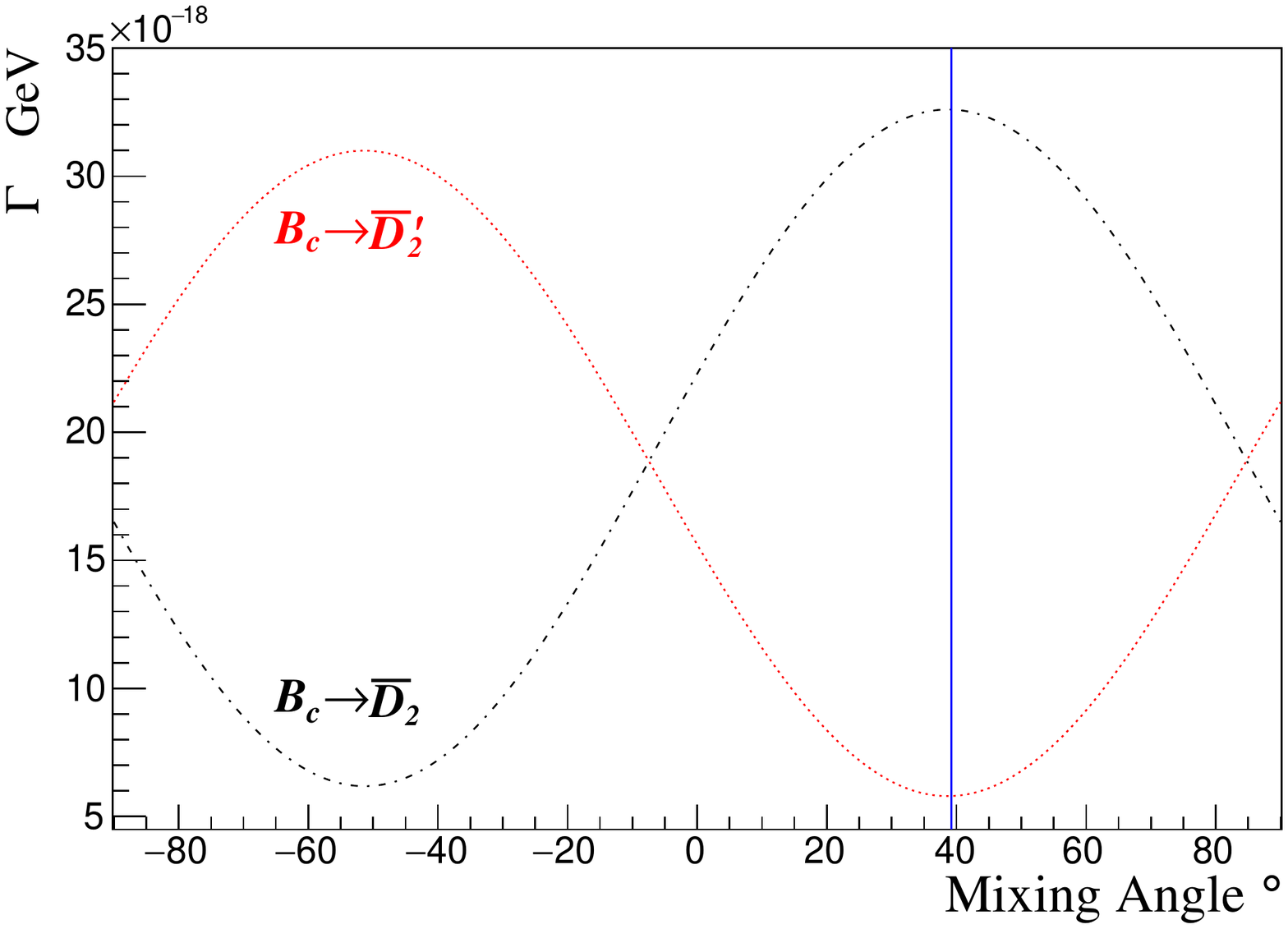} \label{Fig-An1-Bc}}\\
\subfigure[$\Gamma(\bar B\To D_2e\bar \nu)/ \Gamma(\bar B\To D^{\prime}_2e\bar \nu)$ vs mixing angle.]{\includegraphics[width=0.48\textwidth]{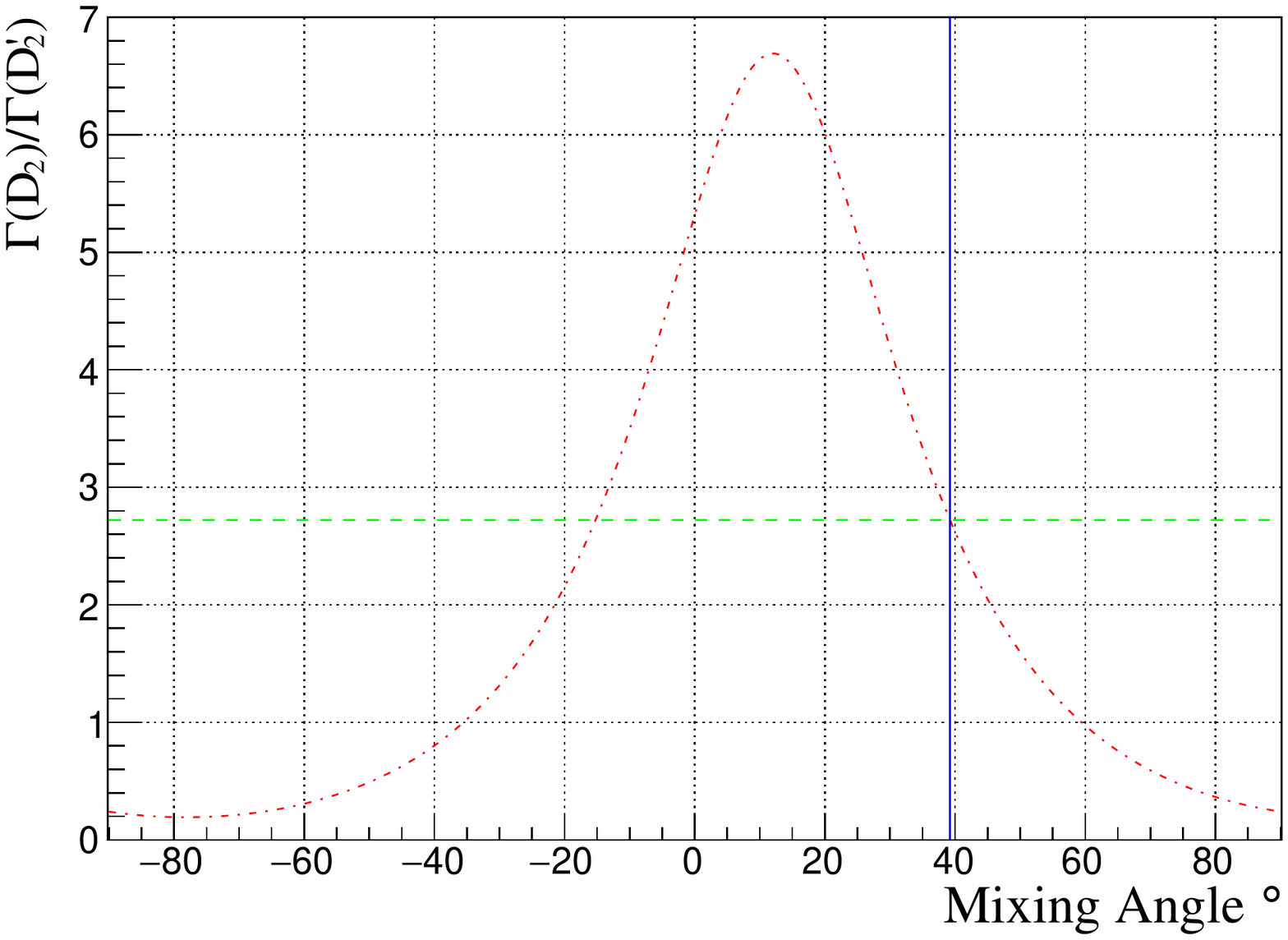} \label{Fig-An2}}
\subfigure[$\Gamma(B^-_c\To \bar D_2e\bar \nu)/\Gamma{(B^-_c\To \bar D^{\prime}_2e\bar \nu)}$ vs mixing angle.]{\includegraphics[width=0.48\textwidth]{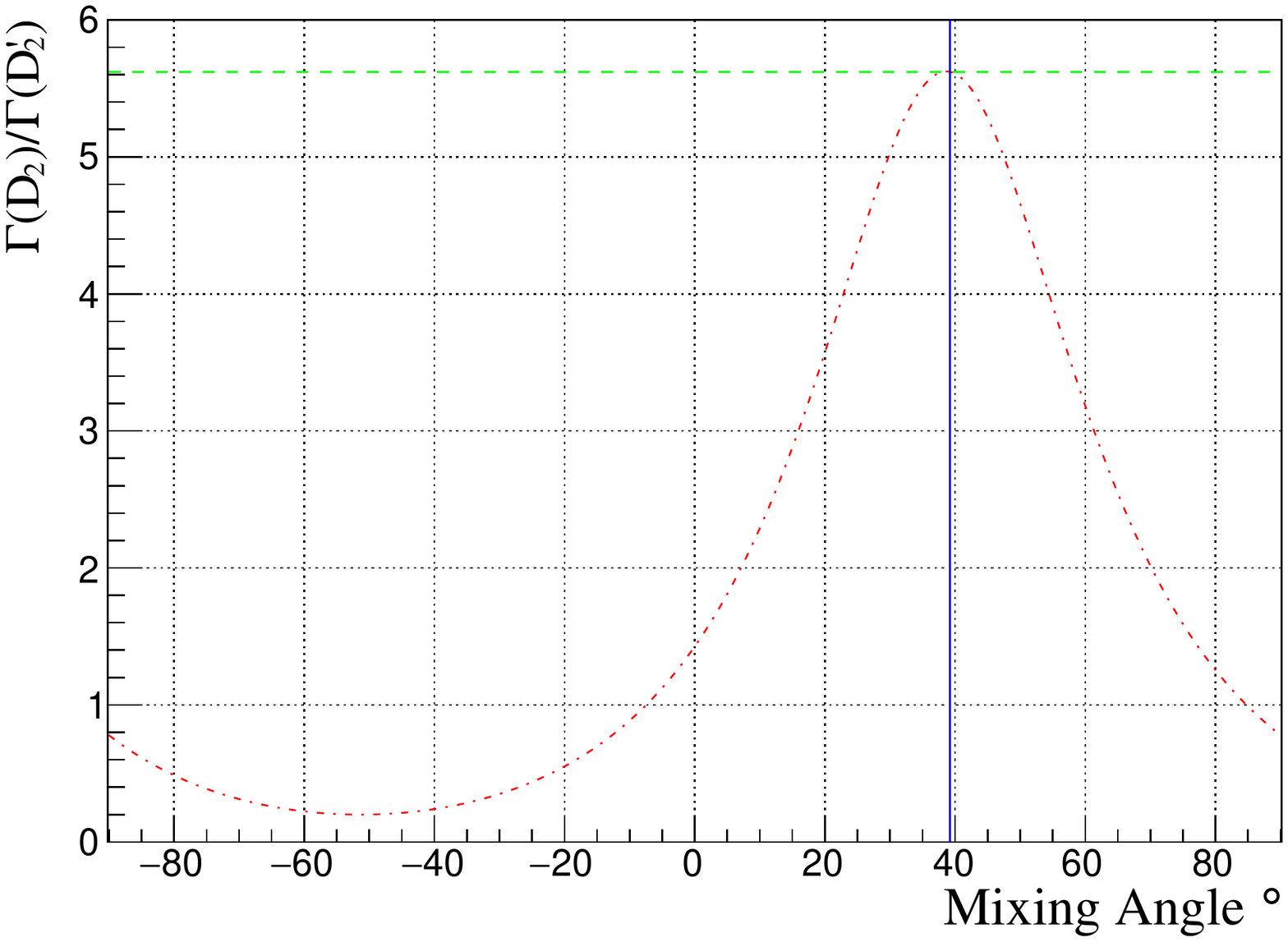} \label{Fig-An2-Bc}}
\caption{Decay widths $\Gamma[\bar B\To D_2(D'_2)e \bar{\nu}]$ and $\Gamma[B^-_c\To\bar D_2(\bar D'_2)e \bar{\nu}]$ vary along with the mixing angle. The vertical solid line shows the results when mixing angle $\alpha=39.23^\circ$, where the decay width ratio is 2.73 for $\bar B\To D_2(D'_2)e \bar{\nu}$ and 5.63 for  $B^-_c\To\bar D_2(\bar D'_2)e \bar{\nu}$.}\label{Fig-width-An}
\end{figure}

The decay widths for $\bar B_{(s)}$ or $B_c$ to $2^-$ states mesons are dependent on the mixing angle $\alpha$, which can be showed by~\autoref{Fig-An1} and \autoref{Fig-An1-Bc}. This dependence for $\bar B$ decays can be described by the following equations
\begin{align}
\Gamma(\bar B\To D_2e\bar\nu) &=\Gamma_1\big[1+\lambda_1\cos (2\alpha+\Theta_1)\big],\label{E-mix1}\\
\Gamma(\bar B\To D'_2e\bar\nu)&=\Gamma_2\big[1-\lambda_2\cos (2\alpha+\Theta_2)\big].\label{E-mix2}
\end{align}
Our fit results give that the parameters are as
\begin{gather*}
\Gamma_1=3.46\times10^{-16},\quad \lambda_1=0.709,\quad \Theta_1=-23.7\degree,\\
\Gamma_2=3.06\times10^{-16},\quad \lambda_2=0.711,\quad \Theta_2=-24.1\degree.
\end{gather*} 
The tiny differences in parameters for $D_2$ and $D'_2$ come from the small difference between $m_{D_2}$ and $m_{D'_2}$. In~\autoref{Fig-An2} and~\autoref{Fig-An2-Bc}, we also show the ratios $\frac{\Gamma(\bar B\to{D_2}e\bar\nu)}{\Gamma(\bar B\to{D'_2}e\bar\nu)}$ and $\frac{\Gamma(B_c^-\to\bar D_2e\bar\nu)}{\Gamma(B^-_c\to\bar D'_2e\bar\nu)}$, which are very sensitive to the mixing angle.

\subsection{Non-leptonic decay widths and branching ratios}
The non-leptonic decay widths are listed in~\autoref{width-non}, where we have kept the Wilson coefficient $a_1$ in order to facilitate comparison with other models. The corresponding branching ratios are listed in~\autoref{br-non}, where we have specified the values $a^b_1=1.14$ for $b\To c(u)$ transition and $a^c_1=1.2$ for $c\To d(s)$ transition~\cite{PRD73-2006}. From the non-leptonic decay results we can see that, with the same final $D$ meson, the $\rho$ mode has the largest branching ratio and can reach $10^{-3}$ order in $\bar B_{(s)}$ decays, and $10^{-6}$ order in $B_c$ decay. When the light mesons have the same quark constituents, the width for decay into vector meson $(\rho, K^*)$ mode is about $2\sim3$ times greater than its pseudoscalar meson $(\pi, K)$ mode.

\begin{table}[ht]
\centering
\caption{Non-leptonic decay widths of $\bar{B}$, $\bar{B}_s$ and $B_c$ to $D$-wave heavy-light meson with general Wilson coefficient $a_1$.}\label{width-non}
\vspace{-0.2em}
\centering
\begin{tabular}{lclclc}
&&&&&\hfill{$\times a_1^2$ [\si{GeV}]}\\
\toprule[2pt]
Channels                     	&  Width               &Channels       		      &Width                &Channels   \hspace{-2cm}	  	  &Width     \\
\midrule[1.3pt]
$\bar{B}\To{D_2} \pi^- $   	&  $2.8\er{0.7}{0.7}{-16} $   & $\bar{B}\To{D'_2} \pi^- $  	 & $9.8\er{2.0}{2.2}{-17} $   & $\bar{B}\To{D^*_3} \pi^- $    & $2.2\er{0.4}{0.4}{-16} $     \\
$\bar{B}\To{D_2} K^-   $   	&  $2.0\er{0.5}{0.5}{-17} $   & $\bar{B}\To{D'_2}  K^-  $  	 & $6.7\er{1.4}{1.5}{-18} $   & $\bar{B}\To{D^*_3} K^-   $	  & $1.4\er{0.3}{0.3}{-17} $   \\
$\bar{B}\To{D_2} \rho^-$   	&  $5.5\er{1.4}{1.4}{-16} $   & $\bar{B}\To{D'_2}\rho^- $  	 & $2.0\er{0.4}{0.4}{-16} $   & $\bar{B}\To{D^*_3} \rho^-$    & $4.7\er{0.8}{0.9}{-16} $     \\
$\bar{B}\To{D_2} K^{*-} $  	&  $2.9\er{0.7}{0.7}{-17} $   & $\bar{B}\To{D'_2}K^{*-} $  	 & $1.0\er{0.2}{0.2}{-17} $   & $\bar{B}\To{D^*_3} K^{*-}$    & $2.5\er{0.4}{0.5}{-17} $    \\
\midrule[1.30pt]
$\bar{B}_s\To{D_{s2}} \pi^- $   &  $4.0\er{1.0}{1.1}{-16} $   & $\bar{B}_s\To{D'_{s2}} \pi^- $   & $1.2\er{0.4}{0.4}{-16} $ & $\bar{B}_s\To{D^*_{s3}} \pi^- $   & $2.9\er{0.8}{0.7}{-16} $ \\
$\bar{B}_s\To{D_{s2}} K^-   $   &  $2.8\er{0.8}{0.7}{-17} $   & $\bar{B}_s\To{D'_{s2}}  K^-  $   & $8.2\er{2.4}{2.5}{-18} $ & $\bar{B}_s\To{D^*_{s3}} K^-   $   & $1.9\er{0.5}{0.5}{-17} $   \\
$\bar{B}_s\To{D_{s2}} \rho^-$   &  $8.1\er{2.3}{2.1}{-16} $   & $\bar{B}_s\To{D'_{s2}}\rho^- $   & $2.4\er{0.7}{0.8}{-16} $ & $\bar{B}_s\To{D^*_{s3}} \rho^-$   & $6.4\er{1.7}{1.7}{-16} $ \\
$\bar{B}_s\To{D_{s2}} K^{*-} $  &  $4.2\er{1.2}{1.1}{-17} $   & $\bar{B}_s\To{D'_{s2}}K^{*-} $   & $1.3\er{0.4}{0.4}{-17} $ & $\bar{B}_s\To{D^*_{s3}} K^{*-}$   & $3.4\er{0.9}{0.9}{-17} $ \\
\midrule[1.3pt]
$B^-_c\To \bar D_2\pi^-$		& $1.1\er{0.4}{0.7}{-18}  $		& $B^-_c\To \bar D'_2\pi^-$	& $2.5\er{0.7}{1.3}{-19}  $ 	& $B^-_c\To \bar D^*_3\pi^-$	& $1.1\er{0.3}{0.5}{-18}  $\\
$B^-_c\To \bar D_2K^-$		& $9.0\er{0.3}{0.5}{-20}  $		& $B^-_c\To \bar D'_2K^-$		& $2.0\er{0.6}{1.0}{-20}  $	& $B^-_c\To \bar D^*_3K^-$		& $8.4\er{2.2}{4.0}{-20}  $\\
$B^-_c\To \bar D_2\rho^-$  	& $3.5\er{1.1}{1.9}{-18}  $ 		& $B^-_c\To \bar D'_2\rho^-$ 	& $7.8\er{2.2}{3.9}{-19}  $	& $B^-_c\To \bar D^*_3\rho^-$	& $3.1\er{0.8}{1.4}{-18}  $\\
$B^-_c\To\bar D_2K^{*-}$ 		& $2.1\er{0.6}{1.1}{-19}  $		& $B^-_c\To\bar D'_2K^{*-}$ 	& $4.7\er{1.3}{2.3}{-20}  $	& $B^-_c\To\bar D^*_3K^{*-}$	& $1.9\er{0.5}{0.8}{-19}  $\\
\midrule[1.3pt]
$B^+_c\To      B_2 \pi^+ $    	& $1.4\er{0.1}{0.1}{-19} $        & $B^+_c\To      B'_2 \pi^+ $   & $1.3\er{0.3}{0.2}{-21} $          & $B^+_c\To      B^*_3 \pi^+ $  	  & $1.9\er{0.4}{0.4}{-21} $     \\
\bottomrule[2pt]
\end{tabular}
\end{table}

\begin{table}[h]
\vspace{0.5cm}
\caption{Branching ratios of non-leptonic decays for $\bar{B}$, $\bar B_s$ and $B_c$ to $D$-wave heavy-light mesons. $a^b_1=1.14$ for $b$ quark decay and $a^c_1=1.2$ for $c$ quark decay.}\label{br-non}
\centering
\begin{tabular}{lclclc}
\toprule[2pt]
       Channels                  &Br~           & Channels                      	& Br     	   	 &       Channels                     & Br   \\
\midrule[1.5pt]
$\bar{B}\To {D_2} \pi^- $   	 & $8.5\er{2.2}{2.1}{-4}$	&$\bar{B}\To {D'_2} \pi^- $   		& $2.9\er{0.6}{0.6}{-4}$   &$\bar{B}\To {D^*_3} \pi^- $         & $6.5\er{1.2}{1.2}{-4}$ \\
$\bar{B}\To {D_2} K^-   $   	 & $5.9\er{1.5}{1.4}{-5}$	&$\bar{B}\To {D'_2}  K^-  $   		& $2.0\er{0.4}{0.4}{-5}$	&$\bar{B}\To {D^*_3} K^-   $         & $4.3\er{0.8}{0.8}{-5}$\\
$\bar{B}\To {D_2} \rho^-$   	 & $1.7\er{0.4}{0.4}{-3}$	&$\bar{B}\To {D'_2}\rho^- $   		& $5.9\er{1.2}{1.3}{-4}$   &$\bar{B}\To {D^*_3} \rho^-$         & $1.4\er{0.3}{0.3}{-3}$ \\
$\bar{B}\To {D_2} K^{*-}$   	 & $8.6\er{2.2}{2.1}{-5}$	&$\bar{B}\To {D'_2}K^{*-}$    		& $3.1\er{0.6}{0.7}{-5}$   &$\bar{B}\To {D^*_3} K^{*-}$         & $7.5\er{1.3}{1.4}{-5}$ \\
\midrule[1.3pt]
$\bar{B}_s\To {D_{s2}} \pi^- $   & $1.2\er{0.3}{0.3}{-3}$	&$\bar{B}_s\To {D'_{s2}} \pi^- $	& $3.6\er{1.0}{1.1}{-4}$   &$\bar{B}_s\To {D^*_{s3}} \pi^- $    & $8.5\er{2.3}{2.2}{-4}$ \\
$\bar{B}_s\To {D_{s2}} K^-   $   & $8.3\er{2.3}{2.1}{-5}$	&$\bar{B}_s\To {D'_{s2}}  K^-  $    	& $2.5\er{0.7}{0.7}{-5}$   &$\bar{B}_s\To {D^*_{s3}} K^-   $    & $5.7\er{1.6}{1.5}{-5}$\\
$\bar{B}_s\To {D_{s2}} \rho^-$   & $2.4\er{0.7}{0.6}{-3}$	&$\bar{B}_s\To {D'_{s2}}\rho^- $    	& $7.3\er{2.1}{2.2}{-4}$   &$\bar{B}_s\To {D^*_{s3}} \rho^-$    & $1.9\er{0.5}{0.5}{-3}$ \\
$\bar{B}_s\To {D_{s2}} K^{*-}$   & $1.3\er{0.4}{0.3}{-4}$	&$\bar{B}_s\To {D'_{s2}}K^{*-}$     	& $3.8\er{1.1}{1.2}{-5}$   &$\bar{B}_s\To {D^*_{s3}} K^{*-}$    & $1.0\er{0.3}{0.3}{-4}$ \\
\midrule[1.3pt]
$B^-_c\To \bar D_2\pi^-$		& $1.0\er{0.3}{0.6}{-6}$	& $B^-_c\To \bar D'_2\pi^-$	& $2.2\er{0.6}{1.2}{-7}$	& $B^-_c\To \bar D^*_3\pi^-$	& $9.6\er{2.6}{4.6}{-7}$ \\
$B^-_c\To \bar D_2K^-$		& $8.0\er{2.5}{4.6}{-8}$	& $B^-_c\To \bar D'_2K^-$		& $1.7\er{0.5}{0.9}{-8}$	& $B^-_c\To \bar D^*_3K^-$		& $7.5\er{0.2}{3.5}{-8}$	\\
$B^-_c\To \bar D_2\rho^-$  	& $3.1\er{1.0}{1.7}{-6}$ 	& $B^-_c\To \bar D'_2\rho^-$ 	& $6.9\er{2.0}{3.5}{-7}$	& $B^-_c\To \bar D^*_3\rho^-$	& $2.8\er{0.7}{1.3}{-6}$ \\
$B^-_c\To\bar D_2K^{*-}$ 		& $1.9\er{0.6}{1.0}{-7}$	& $B^-_c\To\bar D'_2K^{*-}$ 	& $4.2\er{1.2}{2.0}{-8}$	& $B^-_c\To\bar D^*_3K^{*-}$	& $1.7\er{0.4}{0.7}{-7}$ \\
\midrule[1.3pt]
$B^+_c\To      B_2 \pi^+ $    	 & $1.4\er{0.1}{0.1}{-7}$  &$B^+_c\To      B'_2 \pi^+ $   		& $1.3\er{0.2}{0.2}{-9}$    &$B^+_c\To      B^*_3 \pi^+ $   	 & $1.9\er{0.4}{0.4}{-9} $     \\
\bottomrule[2pt]
\end{tabular}
\end{table}
\vspace{1ex}    

\section{Summary}\label{Sec-5}
In this work we calculated semi-leptonic and non-leptonic decays of $\bar B_{(s)}$ into $D$-wave charmed mesons ($D_{(s)2}$, $D'_{(s)2}$, $D^*_{(s)3}$) and $B_c$ into $D$-wave charmed and bottomed excited mesons. Form factors of hadronic transition are calculated by instantaneous Bethe-Salpeter methods. 
The semi-electronic branching ratios for $\bar B_{(s)}\To D^{(*)}_{(s)J}$ we got are about $10^{-3}$ order, and for $B_c$ to $D$-wave charmed mesons are about $10^{-5}$ order. The non-leptonic branching ratios for decays to $\rho$ mode can reach $10^{-3}$ order for $\bar B_{(s)}$ decays. So the $D$-wave $D$ and $D_s$ mesons are hopefully to be detected in $\bar B_{(s)}$ decays by current experiments . Our results reveal the branching fractions for $B_c$ to $D$-wave bottomed mesons are less than $10^{-8}$, which makes the $D$-wave bottomed mesons almost impossible to be discovered in $B_c$ decays by current experiments. 
 
We also present the angular distribution and charged lepton spectra for $\bar B$ and $B_c$ decays. The $2^-$ states $D_2$ and $D'_2$ are the mixing states of $^1\!D_2-{^3\!D_2}$, so we present the dependence of the decay width varying along with the mixing angle. Based on our results, the semi-leptontic and non-leptonic branching ratios for $\bar B_{(s)}$ decays to the $D$-wave charm and charm-strange mesons have reached the experimental detection thresholds. These results would be helpful in future detecting and understanding these new $D$-wave excited $D_{(s)}$ mesons.

\section*{Acknowledgments}
This work was supported in part by the National Natural Science
Foundation of China (NSFC) under Grant Nos.~11405037, 11575048 and 11505039, and in part by PIRS of HIT Nos.~T201405, A201409, and B201506.

\begin{appendix}
\section{Expressions for $N_i$s in the Hadronic Tensor $H_{\mu\nu}$}\label{Ni}

The hadronic tensor $N_i~(i=1,2,4,5,6)$ for $\bar B$ to $D_{2}$ meson are 
\begin{align}
N_1 &= \frac{2 M^4 \bm{p}_F^4 s_1^2}{3 M_F^4}-\frac{4 M^2 \bm{p}_F^2 s_1 s_3}{3 M_F^2}-\frac{1}{2} M^2 \bm{p}_F^2 s_4^2+\frac{s_3^2}{6}, \label{N1-2} \\
N_2 &= \frac{2 E_F M^3 \bm{p}_F^2 s_1 s_3}{3 M_F^4}+\frac{E_F M^3 \bm{p}_F^2 s_4^2}{2 M_F^2}-\frac{E_F M s_3^2}{6 M_F^2}+\frac{2 M^4 \bm{p}_F^4 s_1 s_2}{3 M_F^4}-\frac{2 M^2 \bm{p}_F^2 s_2 s_3}{3 M_F^2},\label{N2-2} \\
N_4 &= \frac{4 E_F M^3 \bm{p}_F^2 s_2 s_3}{3 M_F^4}+\frac{2 M^4 \bm{p}_F^4 s_2^2}{3 M_F^4}-\frac{M^4 \bm{p}_F^2 s_4^2}{2 M_F^2}+\frac{M^2 s_3^2 (M_F^2+4 \bm{p}_F^2)}{6 M_F^4}, \label{N4-2} \\
N_5 &= -\frac{M^4 \bm{p}_F^4 s_4^2}{2 M_F^2}-\frac{M^2 \bm{p}_F^2 s_3^2}{2 M_F^2}, \label{N5-2} \\
N_6 &= -\frac{M^2 \bm{p}_F^2 s_3 s_4}{M_F^2}. \label{N6-2}
\end{align}
Here $\bm p_F$ denotes the three-momentum of final $D$ systems and $E_F=\sqrt{M_F^2+\bm p_F^2}$. For $\bar B$ to $D'_{2}$ the relations between $N_i$ and form factors $t_k~(k=1,2,3,4)$ have the same form with that for $D_2$, just $s_k$ are replaced with $t_k$. Both $s_k$ and $t_k$ are functions of $q'^2_\perp$.

The hadronic tensor $N_i$ for $\bar B$ to $D^*_{3}$ are expressed with form factors $h_k~(k=1,2,3,4)$ as
\begin{align}
N_1 &=  \frac{2 M^6 \bm{p}_F^6 h_1^2}{5 M_F^6}-\frac{4 M^4 \bm{p}_F^4 h_1 h_3}{5 M_F^4}-\frac{4 M^4 \bm{p}_F^4 h_4^2}{15 M_F^2}+\frac{2 M^2 \bm{p}_F^2 h_3^2}{15 M_F^2}, \label{N1-3}\\
N_2 &=  \frac{2 {E_F} M^5 \bm{p}_F^4 h_1 h_3}{5 M_F^6}+\frac{4 {E_F} M^5 \bm{p}_F^4 h_4^2}{15 M_F^4}-\frac{2 {E_F} M^3 \bm{p}_F^2 h_3^2}{15 M_F^4}+\frac{2 M^6 \bm{p}_F^6 h_1 h_2}{5 M_F^6}-\frac{2 M^4 \bm{p}_F^4 h_2 h_3}{5 M_F^4}, \label{N2-3}\\
N_4 &=  \frac{4 {E_F} M^5 \bm{p}_F^4 h_2 h_3}{5 M_F^6}+\frac{2 M^6 \bm{p}_F^6 h_2^2}{5 M_F^6}-\frac{4 M^6 \bm{p}_F^4 h_4^2}{15 M_F^4}+\frac{2 M^4 \bm{p}_F^2 h_3^2 (M_F^2+3 \bm{p}_F^2) }{15 M_F^6}, \label{N4-3}\\
N_5 &=  -\frac{4 M^6 \bm{p}_F^6 h_4^2}{15 M_F^4}-\frac{4 M^4 \bm{p}_F^4 h_3^2}{15 M_F^4}, \label{N5-3}\\
N_6 &=-\frac{8 M^4  \bm{p}_F^4 h_3 h_4}{15 M_{\!F}^4}. \label{N6-3}
\end{align}.

\section{Full Salpeter equations and the numerical solutions}\label{Salpeter-EQ}
\subsection{Salpeter equations}
Salpeter wave function $\varphi(q_\perp)$ is related to BS wave function $\Psi(q)$ by the following definition
\begin{gather}
\varphi(q_\perp)=\text{i}\int \frac{\mathrm{d}q_P}{2\pi}\Psi(q),\quad \eta(q_\perp)=\int \frac{\mathrm{d}^3k_\perp}{(2\pi)^3}\varphi(k_\perp)V(|q_\perp- k_\perp|),
\end{gather}
where the 3-dimensional integration $\eta(q_\perp)$ can be understood as the BS vertex for bound states, and $V(|q_\perp- k_\perp|)$ denotes the instantaneous interaction kernel.

The projection operators $\Lambda^{\pm}_i(q_\perp)$~($i=1$ for quark and 2 for anti-quark) are defined as
\begin{gather}
\Lambda^{\pm}_i=\frac{1}{2\omega_i}\left[ \frac{\slashed P}{M}\omega_i\pm(-1)^{i+1}(m_i+\slashed q_\perp) \right].
\end{gather}
Then we define four wave functions $\varphi^{\pm\pm}$ by $\varphi$ and $\Lambda^\pm_i$ as
\begin{gather}
\varphi^{\pm\pm}\equiv\Lambda_1^\pm(q_\perp)\frac{\slashed P}{M}\varphi(q_\perp)\frac{\slashed P}{M}\Lambda_2^\pm(q_\perp),\label{Def-np}
\end{gather}
where $\varphi^{++}$ and $\varphi^{--}$ are called the positive and negative Salpeter wave function, respectively. And we can easily check that $\varphi=\varphi^{++}+\varphi^{-+}+\varphi^{+-}+\varphi^{--}$.

The full coupled Salpeter equations then can be expressed as~\cite{PR87-1952}
\begin{gather}
\varphi^{+-}=\varphi^{-+}=0\label{BS-np},\\
(M-\omega_1-\omega_2)\varphi^{++}=+\Lambda_1^+(q_\perp)\eta(q_\perp)\Lambda^+_2(q_\perp),\label{BS-pp}\\
(M+\omega_1+\omega_2)\varphi^{--}=-\Lambda_1^-(q_\perp)\eta(q_\perp)\Lambda^-_2(q_\perp).\label{BS-nn}
\end{gather}
From above equations, we can see that in the weak binding condition $M\sim (\omega_1+\omega_2)$, $\varphi^{--}$ is much smaller compared with $\varphi^{++}$ and can be ignored in the calculations. The normalization condition for Salpeter wave function reads
\begin{gather}\label{Norm-BS}
\int \frac{\text{d}^3q_\perp}{(2\uppi)^3}\left[\overline\varphi^{++}\frac{\slashed P}{M}\varphi^{++}\frac{\slashed P}{M}-\overline\varphi^{--}\frac{\slashed P}{M}\varphi^{--}\frac{\slashed P}{M}\right]=2M.
\end{gather}
\subsection{Numerical solutions of $0^-$ state}
Now we take the $0^{-}~(^1\!S_0)$ state as an example to show the details of achieving Sapeter equations' numerical results. The Salpeter wave function for $0^-(^1\!S_0)$ state has the following general form~\cite{PLB584-2004}
\begin{equation} \label{BS-wave-1s0}
\varphi(^1\!S_0)= M\bigg[ k_1 \frac{\slashed P}{M}+k_2 +k_3 \frac{\slashed q_{\perp}}{M}+k_4\frac{\slashed P \slashed q_{\perp}}{M^2}  \bigg ]\gamma^5.
\end{equation}
By utilizing the Eq.~(\ref{BS-np}),  we have the following two constraint conditions
\begin{gather}\label{BS-par-1s0}
k_3= \frac{M(\omega_1-\omega_2)}{m_1\omega_2+m_2\omega_1}k_2,\quad
k_4= -\frac{M(\omega_1+\omega_2)}{m_1\omega_2+m_2\omega_1}k_1.
\end{gather}
In above wave function, the only undetermined wave functions are $k_1$ and $k_2$, which are the functions of $q^2_\perp$.

By using the definition Eq.~(\ref{Def-np}), we can easily get the positive Salpeter wave function of ${^1\!S_0}$ state as Eq.~(\ref{wave-1s0}), and the corresponding constraint conditions Eq.~(\ref{par-1s0}).
Similarly, the Salpeter negative wave function $\varphi^{--}(^1\!S_0)$ is expressed as 
\begin{equation} \label{wave-1s0-nn}
\varphi^{--}(^1\!S_0)=\bigg [ Z_1+Z_2 \frac{\slashed P}{M} +Z_3 \frac{\slashed q_{\perp}}{M}+Z_4\frac{\slashed P \slashed q_{\perp}}{M^2}  \bigg ]\gamma^5.
\end{equation}
$Z_i~(i=1,2,3,4)$ has the following forms
\begin{equation}\label{par-1s0-nn}
\begin{aligned}
Z_1=& \frac{M}{2}\biggl[k_2 -\frac{\omega_1+\omega_2}{m_1+m_2}k_1\bigg],&\quad Z_3=&-\frac{M(\omega_1-\omega_2)}{m_1\omega_2+m_2\omega_1}Z_1,\\
Z_2=&\frac{M}{2}\biggl[k_1-\frac{m_1+m_2}{\omega_1+\omega_2}k_2 \bigg],&\quad Z_4=&+\frac{M(m_1+m_2)}{m_1\omega_2+m_2\omega_1}Z_1.
\end{aligned}
\end{equation}
And now the normalization condition Eq.~(\ref{Norm-BS}) becomes
\begin{gather}\label{Norm-1S0}
\int \frac{\text{d}^3\bm{q}}{(2\uppi)^3}\frac{8M\omega_1\omega_2k_1k_2}{(m_1\omega_2+m_2\omega_1)}=1.
\end{gather}

Inserting the Salpeter positive wave function Eq.~(\ref{wave-1s0}), and negative wave function Eq.~(\ref{wave-1s0-nn}) into Salpeter equations Eq.~(\ref{BS-pp}) and~(\ref{BS-nn}) respectively, we can obtain two coupled eigen equations on $k_1$ and $k_2$~\cite{PLB584-2004} as
\begin{equation}
\left \{
\begin{aligned}
(M-\omega_1-\omega_2)\left[ck_1(\bm q)+k_2(\bm q)\right]=\frac{1}{2\omega_1\omega_2}\int \text{d}^3\bm{k} \left[\beta_{1}k_1(\bm k)+\beta_{2}k_2(\bm k)\right],\\
(M+\omega_1+\omega_2)\left[k_2(\bm q)-ck_1(\bm q)\right]=\frac{1}{2\omega_1\omega_2}\int \text{d}^3\bm{k} \left[\beta_{1}k_1(\bm k)-\beta_{2}k_2(\bm k)\right],
\end{aligned}
\right .
\end{equation}
where we have used definition $c=\frac{\omega_1+\omega_2}{m_1+m_2}$ and the shorthand
\begin{equation}
\begin{aligned}
\beta_1&=\bm k\cdot \bm q(V_s+V_v)\frac{(\nu_1+\nu_2)(\omega_1+\omega_2)}{m_1\nu_2+m_2\nu_1} -(V_s-V_v)(m_1\omega_2+m_2\omega_1),\\
\beta_2&=\bm k\cdot \bm q(V_s+V_v)\frac{(\nu_1-\nu_2)(m_1-m_2)}{m_1\nu_2+m_2\nu_1}           -(V_s-V_v)(m_1m_2+\omega_1\omega_2+\bm q^2).
\end{aligned}
\end{equation}
In above equations, $V_s$ and $V_v$ are the scalar and vector parts defined in Cornell potential~(see Eq.~\ref{Cornell}) respectively; we have used the definition $\nu_i=\sqrt{m^2_i+\bm k^2}~(i=1,2)$.

Then by solving the two coupled eigen equations numerically, we achieve the mass spectrum and corresponding wave functions $k_1$, $k_2$. Repeating the similar procedures we can obtain the numerical wave functions for $^1\!D_2$, $^3\!D_2$ and $^3\!D_3$ states. Interested reader can see more details on solving the full Salpeter equations in Refs.~\cite{3D23-wave,JPG40-2013,PLB584-2004}.

\section{Positive Salpeter wave function for $^1\!S_0$, $^1\!D_2$ and $^3\!D_2$}\label{sec-wave}
The positive Salpeter wave function and its constraint conditions for $^1\!D_2$ state~\cite{JHEP03-2013} are displayed in \ref{wave-1d2} and \ref{par-1d2}. And the undetermined wave function are $f_1$ and $f_2$.
\begin{equation} \label{wave-1d2}
\psi_D(^1\!D_2)=e_{\mu \nu} q'^\mu_{\perp} q'^\nu_{\perp} \bigg [ b_1+b_2 \frac{\slashed P_F}{M_F} +b_3 \frac{\slashed q'_\perp}{M_F}+b_4\frac{\slashed P_F \slashed q'_{\perp}}{M_F^2} \bigg]\gamma^5,
\end{equation}
\begin{equation}\label{par-1d2}
\begin{aligned}
b_1=& \frac{1}{2}\Big[f_1+\frac{\omega'_1+\omega'_2}{m'_1+m'_2}f_2 \Big],~
&b_3=& -\frac{M_F(\omega'_1-\omega'_2)}{m'_1\omega'_2+m'_2\omega'_1}b_1,\\
b_2=& \frac{1}{2}\Big[f_2+\frac{m'_1+m'_2}{\omega'_1+\omega'_2}f_1 \Big],~
&b_4=& -\frac{M_F(m'_1+m'_2)}{m'_1\omega'_2+m'_2\omega'_1}b_1 .
\end{aligned}
\end{equation}

The positive Salpeter wave function of ${^3\!D_2}$ state~\cite{3D23-wave} and constraint conditions can be written as
\begin{equation} \label{wave-3d2}
\psi_D(^3\!D_2)=\mathrm{i}\epsilon_{\mu \nu \alpha \beta }\frac{P_F^{\nu}}{M_F}q'^{\alpha}_{\perp} e^{\beta \delta} q'_{\perp \delta} \gamma^{\mu} \bigg[ i_1+i_2 \frac{\slashed P_F}{M_F} +i_3 \frac{\slashed q'_\perp}{M_F}+i_4\frac{\slashed P_F \slashed q'_{\perp}}{M_F^2}  \bigg ].
\end{equation}
\begin{equation}\label{par-3d2}
\begin{aligned}
i_1=& \frac{1}{2} \biggl[v_1-\frac{\omega'_1+\omega'_2}{m'_1+m'_2}v_2  \bigg],~
&i_3=&+\frac{M_F(\omega'_1-\omega'_2)}{m'_1\omega'_2+m'_2\omega'_1}i_1,\\
i_2=& \frac{1}{2} \biggl[v_2-\frac{m'_1+m'_2}{\omega'_1+\omega'_2}v_1  \bigg],~
&i_4=&-\frac{M_F(m'_1+m'_2)}{m'_1\omega'_2+m'_2\omega'_1}i_1.
\end{aligned}
\end{equation}
Here we also only have two undetermined wave function $v_1$ and $v_2$.

In above equations \ref{wave-1d2}~$\sim$~\ref{par-3d2} the indeterminate wave functions, such as $f_1$ and $f_2$ in $\psi_D(^1\!D_2)$, $v_1$ and $v_2$ in $\psi_D(^3\!D_2)$, which are functions of $q'^2_\perp$ and can be determined numerically by solving the coupled Salpeter eigen equations \ref{BS-pp} and \ref{BS-nn}.

\end{appendix}

\end{document}